\documentclass[sigconf,screen,nonacm]{acmart}

\graphicspath{{figures/}}

\AtBeginDocument{%
  }

\usepackage{xcolor}

\title[\sys]{\sys: Modifying Applications with Computer Use Agents and Generative UI}

\author{Leping Qiu}
\affiliation{%
  \institution{Autodesk Research}
  \city{Toronto}
  \country{Canada}
}
\email{leping@dgp.toronto.edu}

\author{Jo Vermeulen}
\affiliation{%
  \institution{Autodesk Research}
  \city{Toronto}
  \country{Canada}
}
\email{jo.vermeulen@autodesk.com}

\author{Tim Zindulka}
\affiliation{%
  \institution{Autodesk Research}
  \city{Toronto}
  \country{Canada}
}
\email{tim.zindulka@uni-bayreuth.de}

\author{George Fitzmaurice}
\affiliation{%
  \institution{Autodesk Research}
  \city{Toronto}
  \country{Canada}}
\email{george.fitzmaurice@autodesk.com}

\author{Fraser Anderson}
\affiliation{%
  \institution{Autodesk Research}
  \city{Toronto}
  \country{Canada}}
\email{fraser.anderson@autodesk.com}

\renewcommand{\shortauthors}{Qiu et al.}
\newcommand{\sys}{Interface Quilt}

\begin{abstract}

Modifying an existing application is difficult when it does not provide programmatic access for that change. Recent advances in computer-use agents (CUAs) enable richer understanding and operation of existing interfaces, while generative UI (GenUI) makes it practical to generate new ones. We examine how these advances expand software modification through a design space of what can be changed and a framework for how modifications can be implemented. We then introduce Interface Quilt, a system that combines CUAs and GenUI to generate interface patches, which are new interfaces displayed over an existing application to modify its presentation, interaction, and capabilities. Through demonstrations across a variety of applications, we show that interface patches support a range of modifications without source code, APIs, or accessibility data.

\end{abstract}




\begin{document}

\begin{teaserfigure}
  \includegraphics[width=\textwidth]{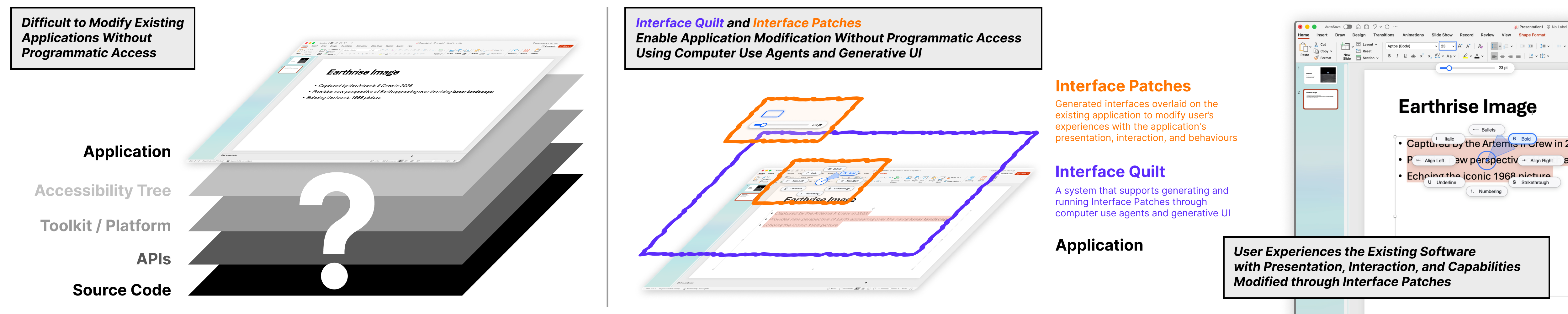}
  \caption{Visual abstract summarizing Interface Quilt -- A system using computer use agents and generative UI to generate and run interface patches that modify the user's experience with the target application's presentation, interaction, and capabilities.}
  \label{fig:teaser}
\end{teaserfigure}

\maketitle

\section{Introduction}
\label{sec:introduction}

Software today is typically designed to be able to satisfy most users' needs most of the time, but it rarely fits how all users work all of the time. One may want to hide or surface controls based on their tasks, replace one interaction technique with another, or make use of the powerful capabilities locked away by professional software without learning the complex interface. This vision of software reshaped for individual users has long been a pursuit of HCI research ~\cite{kayPersonalDynamicMedia1977,stallmanEMACSExtensibleCustomizable1981,macleanUsertailorableSystemsPressing1990,hendersonTheresNoPlace2020}. Prior approaches to achieving this were through customization and tailoring~\cite{macleanUsertailorableSystemsPressing1990,pageUserCustomizationWord1996,myersScriptingGraphicalApplications1998,cookAppleScript2007}, adaptive and adaptable interfaces~\cite{findlaterComparisonStaticAdaptive2004,gajosSUPPLEAutomaticallyGenerating2004,todiAdaptingUserInterfaces2021}, end-user programming and development~\cite{nardiSmallMatterProgramming1993,liebermanEndUserDevelopment2006,koStateArtEnduser2011}, meta-design~\cite{fischerMetadesignManifestoEnduser2004}, and malleable software~\cite{klokmoseWebstratesShareableDynamic2015,radleCodestratesLiterateComputing2017,borowskiVarvReprogrammableInteractive2022,MalleableSoftwareRestoring2025,beaudouin-lafonApplicationsInteractionSubstrates2023,mackayInteractionSubstratesCombining2025}.
In this work, we use \textit{modification} broadly to describe any departure from the software shipped, including changes to what a user perceives (presentation), how they interact (interaction), or what they can do with it (capability). The term follows communities such as game modders and hardware hackers, who alter, extend and retrofit artifacts beyond what their maker anticipated~\cite{thielMalleableGamesLiterature2019, paulGameChangersExploring2026, tanenbaumDemocratizingTechnologyPleasure2013, hurstEmpoweringIndividualsDoityourself2011, ramakersRetroFabDesignTool2016}, as well as designers who repurpose existing applications and devices as prototyping tools for new behaviours~\cite{hartmannReflectivePhysicalPrototyping2006, ledoAstralPrototypingMobile2019}.

Prior modification efforts largely rely on adapting the software through programmatic access pathways provided by the developer, such as scripts, plugins, APIs, or source code. These pathways can limit the extent to which a modification can be applied.  Modifying an application without developer support can be more difficult, but it is not impossible. Prior HCI systems have used pathways not designed for modification, ranging from websites' DOM~\cite{bolinAutomationCustomizationRendered2005,nicholsHighlightSystemCreating2008,zhangFusionOpportunisticWeb2018}, operating system (OS) window APIs~\cite{stuerzlingerUserInterfaceFacades2006, tanWinCutsManipulatingArbitrary2004}, accessibility APIs~\cite{zhangInteractionProxiesRuntime2017,liSUGILITECreatingMultimodal2017,luInteractOutLeveragingInteraction2024}, to runtime code injection to specific application frameworks~\cite{eaganCrackingCocoaNut2011,mengWADESimplifiedGUI2014}. However, not all applications provide these pathways, and there have been various systems that use the graphical user interface (GUI) directly and pixel-based techniques to modify applications. Systems such as \textit{Prefab} add advanced interaction techniques by identifying common UI elements ~\cite{dixonPrefabImplementingAdvanced2010,dixonContentHierarchyPixelbased2011,dixonPixelbasedReverseEngineering2013} and Sikuli provided interface scripting through icons~\cite{yehSikuliUsingGUI2009}. These systems demonstrate that interface structure and actions can be recovered without source-code access. However, these techniques require custom-authored models and templates targeting specific UI toolkits, limiting their applicability.

Recent advances in UI understanding~\cite{zhangScreenRecognitionCreating2021,wuScreenParsingReverse2021,luOmniParserPureVision2024,youFerretUIGroundedMobile2024} and computer-use agents (CUAs)~\cite{agasheAgentOpenAgentic2024,zhangUFO2DesktopAgentOS2025,qinUITARSPioneeringAutomated2025} make it possible to recover richer structure and semantics from rendered interfaces and act through them. At the same time, advances in generative UI (GenUI)~\cite{wuUICoderFinetuningLarge2024,chenGenerativeInterfacesLanguage2025,leviathanGenerativeUILLMs2026,caoGenerativeMalleableUser2025,chenGenUIStudyExploring2025} make it increasingly practical to generate new interactive surfaces. To understand how these capabilities expand software modification, we first develop a design space that characterizes what a modification can change, including changes enabled by CUAs and GenUI to how users perceive and interact with existing software. We then develop a framework for how modifications are achieved, characterizing both the access a modifying system has to an application and the level of synchronization between the modifying system and the target application. Together, the design space and framework help us examine what is possible when modifying applications that reveal rendered pixel data alone.

We introduce \sys{}, a system for modifying existing applications by generating and running \textit{interface patches}: generated interfaces that change an application's presentation, interaction, or capabilities while leaving the application itself unchanged. At build time, \sys{} provides a harness for computer-use agents to explore the target application, inspect its interface state, and discover the controls needed for a modification. \sys{} records this recovered information and makes it available to coding agents, which use it to generate interface patches. An interface patch modifies presentation by covering or augmenting parts of the existing interface, interaction by mediating application output and user input, and capabilities through new logic packaged in the interface patch. At runtime, \sys{} orchestrates interface patches, continuously synchronizing application state to interface patches and translating the user's interactions with the interface patches back into actions on the application. This synchronization makes the interface patch and underlying application behave as a single integrated interface from the user's perspective, creating the experience of modified software without programmatic access to the target application.

We examine \sys{}'s capabilities and boundaries through demonstrations, following Ledo et al.~\cite{ledoEvaluationStrategiesHCI2018}. Across PowerPoint, Affinity, Fusion, Slack, and Overleaf, we use \sys{} to create interface patches inspired by classical interaction techniques such as put-that-there~\cite{boltPutthatthereVoiceGesture1980}, marking menus~\cite{kurtenbachIssuesCombiningMarking1991}, Toolglass~\cite{bierToolglassMagicLenses1993}, Side Views~\cite{terrySideViewsPersistent2002}, Patina~\cite{matejkaPatinaDynamicHeatmaps2013}, Ephemeral Adaptation~\cite{findlaterEphemeralAdaptationUse2009}, SILK~\cite{landaySILKSketchingInterfaces1996}, and ToolTrack~\cite{massonSuperchargingTrialandErrorLearning2022}. We build interface patches that recompose task-centric interfaces, restructure interface controls, add in-situ support, and enhance software capabilities. Together, these demonstrations showed that when the states and controls needed by a modification are visible through the interface, \sys{} can restructure presentation, introduce new interaction techniques, and add capabilities while managing interface patches synchronized with the target application.

This paper makes three contributions. (1) we present \sys{}, a system architecture for generating and running interface patches that modify existing applications through their rendered interfaces; (2) through demonstrations across applications, we show how interface patches can modify presentation, interaction, and capabilities; and (3) we present a design space of software modification, examining changes described in prior systems, and a software framework that separate what a modification changes from how it is achieved, and situate GenUI and CUAs both as forms of modification and as mechanisms for realizing modifications.

\section{Related Work}
\label{sec:related-work}

This work builds on prior work in end-user customization and software modification and is enabled by the emerging capabilities of generative AI models.

\subsection{Vision of Modifiable Software}

Software that can be reshaped by the individual user has long been a goal of investigation by researchers in interactive computing. In Personal Dynamic Media, Kay and Goldberg described Smalltalk that enables any user to mould by their own needs~\cite{kayPersonalDynamicMedia1977}. Stallman built the Emacs text editor with extensibility principles that ``users are not limited by the decisions made by the implementers''~\cite{stallmanEMACSExtensibleCustomizable1981}. MacLean et al.\ argued that no single design can suit all users and situations, thus systems should be tailorable for users' modification~\cite{macleanUsertailorableSystemsPressing1990}. Henderson and Kyng similarly described continuous design in use, where software systems prepare for user modification in response to users' idiosyncratic use and the complexity of the world not anticipated by designers~\cite{hendersonTheresNoPlace2020}.

HCI research demonstrates benefits in modifying applications to better fit the user's tasks, computing environments and abilities. For tasks, users customize a feature-rich word processor to reach frequently used functions faster~\cite{pageUserCustomizationWord1996}, and users are provided with a personal interface with only the functions they chose, navigated and learned better with more sense of control~\cite{mcgrenereEvaluationMultipleInterface2002}. For computing environment change, interfaces regenerated for a device's input and output constraints, such as screen sizes, allow applications in service of context that the designers did not build for~\cite{theveninPlasticityUserInterfaces1999, calvaryUnifyingReferenceFramework2003, gajosSUPPLEAutomaticallyGenerating2004, nicholsMobilizationDemonstrationUsing2008}. For abilities, the modifying offers the most significant benefit, as interfaces generated from a model of a user's motor abilities allow them to complete tasks that they could not perform in the original interface~\cite{gajosAutomaticallyGeneratingUser2007, gajosImprovingPerformanceMotorimpaired2008, wobbrockAbilityBasedDesignConcept2011, zhangInteractionProxiesRuntime2017, zhangScreenRecognitionCreating2021}.

To support modifications, HCI uses more specific terms across different dimensions. \textit{Adaptive} interfaces are changed by the system, whereas \textit{adaptable} interfaces are changed by the user~\cite{findlaterComparisonStaticAdaptive2004}. \textit{End-user programming} and \textit{end-user development} identify users who create computational behaviour without professional programming experiences~\cite{nardiSmallMatterProgramming1993, liebermanEndUserDevelopment2006, koStateArtEnduser2011}. M\o{}rch's customization, integration and extension dimensions distinguish how deeply a change reaches into the software~\cite{morchThreeLevelsEndUser1997}.

Other terms distinguish how modification is made. \textit{Customization} and \textit{tailoring} use mechanisms deliberately provided by the developer, such as preference dialog and scripting languages~\cite{macleanUsertailorableSystemsPressing1990, pageUserCustomizationWord1996, myersScriptingGraphicalApplications1998, cookAppleScript2007}. Plugin extension APIs are similarly developer-supplied contacts for modifications. \textit{Meta-design} withholds the specific design at the time for users to complete during use~\cite{fischerMetadesignManifestoEnduser2004}. \textit{Malleable software} goes further and makes the data and code substrates available for change. Systems like \textit{Webstrates}~\cite{klokmoseWebstratesShareableDynamic2015}, \textit{Codestrates}~\cite{radleCodestratesLiterateComputing2017}, and \textit{Varv}~\cite{borowskiVarvReprogrammableInteractive2022} treat applications and documents as the same modifiable material and interaction substrates decompose applications into substrates and instruments in service of modifications~\cite{MalleableSoftwareRestoring2025, beaudouin-lafonApplicationsInteractionSubstrates2023, mackayInteractionSubstratesCombining2025}. For our purpose, these concepts distinguish the routes to modification rather than the act of modification itself, and are all considered as modifications.

These modification pathways share an assumption that software developers prepare modification mechanisms from within the software, such as packaged options in customization, predefined paths in adaptive software, or software with malleability designed from the ground up. Without this support, modifications become much more difficult~\cite{borowskiVarvReprogrammableInteractive2022}. Our work focuses on providing \sys{} that sits over existing software to provide the user with the experience of modified software, even without providing mechanisms for modification.

\subsection{Modifying Application Without Support Made for Modifications}

It is possible to modify software without support or access built into the software. Prior work has introduced pathways from outside the application for modifications, such as through web code, accessibility APIs, language-specific hooks, and operating systems' APIs.

Web applications are the most readily modified applications, as web pages are declarative documents in HTML and DOMs with client-side source code inspectable for a modifying system. For example, \textit{Chickenfoot} creates automation and customization abilities for end-users by keyword matching the source code ~\cite{bolinAutomationCustomizationRendered2005}. \textit{Highlight} reauthors an existing website for mobile phone and omits pages not used during user demonstrations~\cite{nicholsHighlightSystemCreating2008}. Other systems record and share web operation procedures, inject widgets, spreadsheets, and panels to web pages that were not designed to host them~\cite{leshedCoScripterAutomatingSharing2008, toomimAttachingUIEnhancements2009, littWildcardSpreadsheetdrivenCustomization2020, zhangFusionOpportunisticWeb2018, kimStyletteStylingWeb2022, haoChatClicksGUI2026, luAwarenessActionExploring2024}. The access to client-side code, combined with the ability to examine function calls, provides deeper modifications. For example, \textit{d.mix} enables flexible re-authoring and mashups of API-based web applications~\cite{hartmannProgrammingSampleRapidly2007} and \textit{Genie} reverse-engineers a web application's commands to retarget them as speech, keyboard and command line interface~\cite{swearnginGenieInputRetargeting2017}.

However, many non-web applications on desktop and mobile lack direct source code access and rely on alternative pathways. Some application frameworks support mechanisms for runtime code-injection that systems can reverse-engineer for modification. For example, \textit{Scotty} proposed a runtime toolkit overloading to change macOS Cocoa applications' method implementations at runtime~\cite{eaganCrackingCocoaNut2011}, \textit{WADE} uses Windows' dynamic linking library to modify interface layout and behaviours~\cite{mengWADESimplifiedGUI2014}, and earlier toolkit research taps into the operating system's drawing pipelines to transform and extend interface widgets~\cite{edwardsSystematicOutputModification1997, olsenImplementingInterfaceAttachments1999}. Some operating systems, such as Windows and Android, provide a windowing system with region names and input redirection, useful for basic layout reconfigurations. \textit{WinCuts} use them to duplicate live regions of arbitrary windows~\cite{tanWinCutsManipulatingArbitrary2004}, \textit{User Interface Facades} to recombine and replace widgets of running applications~\cite{stuerzlingerUserInterfaceFacades2006}, and \textit{A-Mash} extracts parts of unmodified Android applications into a single wrapper application~\cite{leeAmashProvidingSingleapp2022}. Specifically for input redirection, \textit{ICON} reroutes device inputs with Windows APIs~\cite{dragicevicInputConfiguratorToolkit2004} and \textit{HIDAgent} simulate human-interface device controls through an external hardware~\cite{bighamHIDAgentToolkitEnabling2026}. The accessibility API offers UI element trees with markups designed for screen readers, but it also enables the ``curb cut effect'' for modification systems to structurally understand the interface. For example, \textit{Interaction Proxies} use it to provide augmented metadata and interaction mediation for runtime accessibility repair on Android~\cite{zhangInteractionProxiesRuntime2017}. The API provides abilities to record user demonstration, input manipulation, and interface augmentation~\cite{liSUGILITECreatingMultimodal2017,liPUMICEMultiModalAgent2019,luInteractOutLeveragingInteraction2024,huangInteractionProxyManager2023, matejkaPatinaDynamicHeatmaps2013}. Yet, it still relies on developers to provide proper support, which is not always guaranteed~\cite{zhangInteractionProxiesRuntime2017, changAssociatingVisualRepresentation2011}. Although these pathways allow access to modify some application behaviours, they are limited by the toolkits, platforms and developer support.

The most common denominator shared across GUI applications is the rendered pixels. As such, researchers investigated pixel-based reverse engineering techniques for modifying an application. \textit{Prefab} identifies widget location and outline by template-matching the pixels of known widgets, then retrofits advanced interaction techniques onto unmodified applications~\cite{dixonPrefabImplementingAdvanced2010, dixonContentHierarchyPixelbased2011, dixonGeneralpurposeTargetawarePointing2012, dixonPixelbasedReverseEngineering2013}. Similarly, \textit{Sikuli} used screenshots templates and synthetic inputs to automate desktop workflows~\cite{yehSikuliUsingGUI2009}, and \textit{Waken} recovers widgets and usage from tutorial videos~\cite{banovicWakenReverseEngineering2012}. However, pixel-based methods also introduce limitations and constraints to the modification. First, these methods lack state awareness of the applications. To mitigate this, \textit{Prefab} reintroduce accessibility APIs to mitigate applications' state and style ~\cite{dixonPixelbasedMethodsWidget2014, dixonPrefabLayersPrefab2014}. Second, these methods rely on hand-crafted templates or fixed widget styles from a standard UI toolkit, limiting the general applicability.

Modifying an application without support, therefore, remains limited in the extent of supported applications and modifications. In \sys{}, we extend the pixel-based technique in \textit{Prefab}~\cite{dixonPrefabLayersPrefab2014} and recontextualize the layered interface architecture in \textit{Interaction Proxies}~\cite{zhangInteractionProxiesRuntime2017} with the latest development in generative UI for creating the modifications and computer use agents to explore existing software and bind modifications to activate them.

\subsection{Emerging Capabilities: Generative UI and Computer Use Agents}

Recent developments in generative UI (GenUI) and Computer Use Agents (CUAs), driven by large language models (LLMs) and visual language models (VLMs), lower the barriers to reproducing interfaces and actions. These capabilities create new opportunities for modifying existing applications, although GenUI and CUAs alone have largely been developed for different ends.

Prior to LLMs, user interface management systems and model-based methods have generated interfaces from domain models in order to design once and target any device~\cite{wiechaITSToolRapidly1990, vanderdoncktEncapsulatingKnowledgeIntelligent1993, puertaModelbasedInterfaceDevelopment1997, moriDesignDevelopmentMultidevice2004, limbourgUSIXMLLanguageSupporting2005}. \textit{Graceful Degradation} investigates the transformation rules of interfaces while preserving continuity across platforms. Systems like \textit{Personal Universal Controller} and \textit{Huddle} generate a more usable interface for appliances~\cite{nicholsGeneratingRemoteControl2002, nicholsHuddleAutomaticallyGenerating2006} and \textit{SUPPLE} render an optimized interface based on device and user model~\cite{gajosSUPPLEAutomaticallyGenerating2004}. Other work like \textit{Bespoke} investigated synthesizing a custom UI from the command line~\cite {vaithilingamBespokeInteractivelySynthesizing2019}.

The emerging wave of GenUI takes advantage of LLMs' coding capabilities and focuses on producing usable interfaces from scratch~\cite{wuUICoderFinetuningLarge2024, chenGenerativeInterfacesLanguage2025, leviathanGenerativeUILLMs2026}. GenUI can refer to both runtime generations and adaptations or design-time generations~\cite{chenGenUIStudyExploring2025}. To constrain the generation process, systems like \textit{BISCUIT}~\cite{chengBISCUITScaffoldingLLMGenerated2024} and \textit{DynaVis}~\cite{vaithilingamDynaVisDynamicallySynthesized2024} scaffold widget generation with access to source code, \textit{SQUIRE}~\cite{leungSQUIREInteractiveUI2025} and \textit{Athena}~\cite{beasonAthenaIntermediateRepresentations2026} reintroduce the intermediate models, \textit{Jelly}~\cite{caoGenerativeMalleableUser2025} created data models and UI specifications for gradual UI generation, \textit{Maru}~\cite{kimMaruInformationArchitecture2026} created information architecture to stabilize generation across iterations. Recent work on revibing HCI interactive systems with coding agents suggests coding agents are becoming increasingly successful at producing interfaces~\cite{adarRevibingCodePapers2026}. However, these GenUIs are all built to code bases that the GenUI system owns, and it remains difficult to integrate GenUI into existing applications~\cite{caoGenerativeMalleableUser2025, chenGenerativeInterfacesLanguage2025, xuDuetUIBidirectionalContext2026}. Systems that integrate into existing applications are done through APIs~\cite{liuTaskLensGeneratingTaskConditioned2026} or via the web~\cite{haoChatClicksGUI2026}.

UI understanding and CUAs have developed together. Development in screen recognition enabled systems to recover interface semantics~\cite{zhangScreenRecognitionCreating2021, wangScreen2WordsAutomaticMobile2021}, UI models~\cite{wuScreenParsingReverse2021}, interface texts, layouts, and widgets~\cite{luOmniParserPureVision2024, youFerretUIGroundedMobile2024, akoucheTargetFinderDetectingWidgets2026} from screenshots alone. The understanding feeds in for computer use agents to operate applications on behalf of the users~\cite{agasheAgentOpenAgentic2024, zhangUFO2DesktopAgentOS2025, qinUITARSPioneeringAutomated2025}. Some systems also fuse screenshots with accessibility API or DOM when available to improve performance~\cite{zhangUFO2DesktopAgentOS2025, zhangAPIAgentsVs2025}. To promote users' awareness, control, and learnability when delegating tasks to CUA, researchers have explored adding pausing, checking, and handoff mechanisms~\cite{pengMoraeProactivelyPausing2025, mozannarMagenticUIHumanintheloopAgentic2025, wuVeriOSQueryDrivenProactive2026, changSidekickDesigningCommunication2026, khuranaItMeVs2025}. To improve CUA performance both in accuracy and speed, researchers explored separating build time from runtime process, where a system explores the application once, then packages the recovered controls as a deterministic tool, such as model context protocol (MCPs), for another agent to call on at runtime~\cite{luAXISEfficientHumanAgentComputer2025, zhongActionEngineReactiveProgrammatic2026, wangImperativeDeclarativeLLMfriendly2026, qiuCI4ASemanticComponent2026, liALLOYGeneratingReusable2025}.

We combine the capabilities of GenUI and CUA in \sys{} to create a persistent interaction instrument~\cite{beaudouin-lafonInstrumentalInteractionInteraction2000} as interface patches bounded over an existing application. \sys{} uses CUAs to recover the controls over existing applications and uses coding agents and GenUIs to construct an interface layer for modification. We view these capabilities as emerging mechanisms in a broader space of software modification (\autoref{sec:concept}), enabling changes to how users perceive and interact with software that was not designed for modification.

\subsection{Prior Design Spaces for Software Modification}

Prior work has developed design spaces for specific forms of software modification. \textit{Interaction Proxies} provided interaction re-mappings for runtime accessibility repairs~\cite{zhangInteractionProxiesRuntime2017}, and \textit{InteractOut} extends the space to input manipulations for curbing phone overuse~\cite{luInteractOutLeveragingInteraction2024}. \textit{DOMSteer} describes three operations for in-situ assistance manipulating websites: insert, mutate and recompose~\cite{haoChatClicksGUI2026}. Design spaces for adaptive interfaces similarly describe system-initiated changes to presentation~\cite{gajosExploringDesignSpace2006, findlaterDesignSpaceEvaluation2009}. Zindulka et al.\ develops design space for touch-based LLM interactions by additive, subtractive, and transformative change to text~\cite{zindulkaExploringMobileTouch2025}. Min et al.\ characterize malleable overview-detail interfaces by content, composition and layout~\cite{minMalleableOverviewDetailInterfaces2025, minMeridianDesignFramework2025}, and create a design space of abstraction techniques covering unit of interest, granularity, representation, transformability, presentation and guidance~\cite{minMakingAbstractionConcrete2026}.

These design spaces are scoped to specific mechanisms, interface forms or domains. Our goal is to describe modifications across them by separating what changes for the user and how the change is achieved. The design space characterizes modifications along five dimensions of the user's perception-action loop, including the element and structure of presentation, the modality and steps of interaction, and the capabilities. It further distinguishes reductive, replacement, and additive changes, adapting from the distinction of the modified target used by Zindulka et al.\ for text manipulation~\cite{zindulkaExploringMobileTouch2025}. The framework of modification mechanisms describes how a modification receives and returns states and actions with an application, extending Daniel et al.'s degrees of interoperability~\cite{danielUnderstandingUIIntegration2007}. To our knowledge, our design space and framework are the first to view GenUI and CUAs as both forms of modification and pathways through which other modifications can be achieved.

\section{Design Space and Framework}
\label{sec:concept}

We provide a unified account of software modification by distinguishing \textit{what changes for the user} from \textit{how that change is achieved}. We develop a design space that characterizes the effects of modification on the user's experience of an existing application, and a complementary framework of modification mechanisms that characterizes how modifying systems gain access to the application and remain connected to its state and actions. We derive both by synthesizing the HCI systems reviewed in our related work and drawing on prior design spaces of customization, adaptation, and modification.

We define an \textit{existing application} as an interactive application packaged for the end user with its interface and behaviour fixed by its developers, and refer to this shipped state as the \textit{baseline}. A \textit{modification} is any departure from this baseline that changes what the user perceives, how they interact with the application, or what they can do with it. Using the baseline as a common reference point allows us to compare modifications independently of the mechanisms used to implement them.

\subsection{Design Space of Software Modification}
\label{sec:design-space}

\begin{figure}
    \centering
    \includegraphics[width=\linewidth]{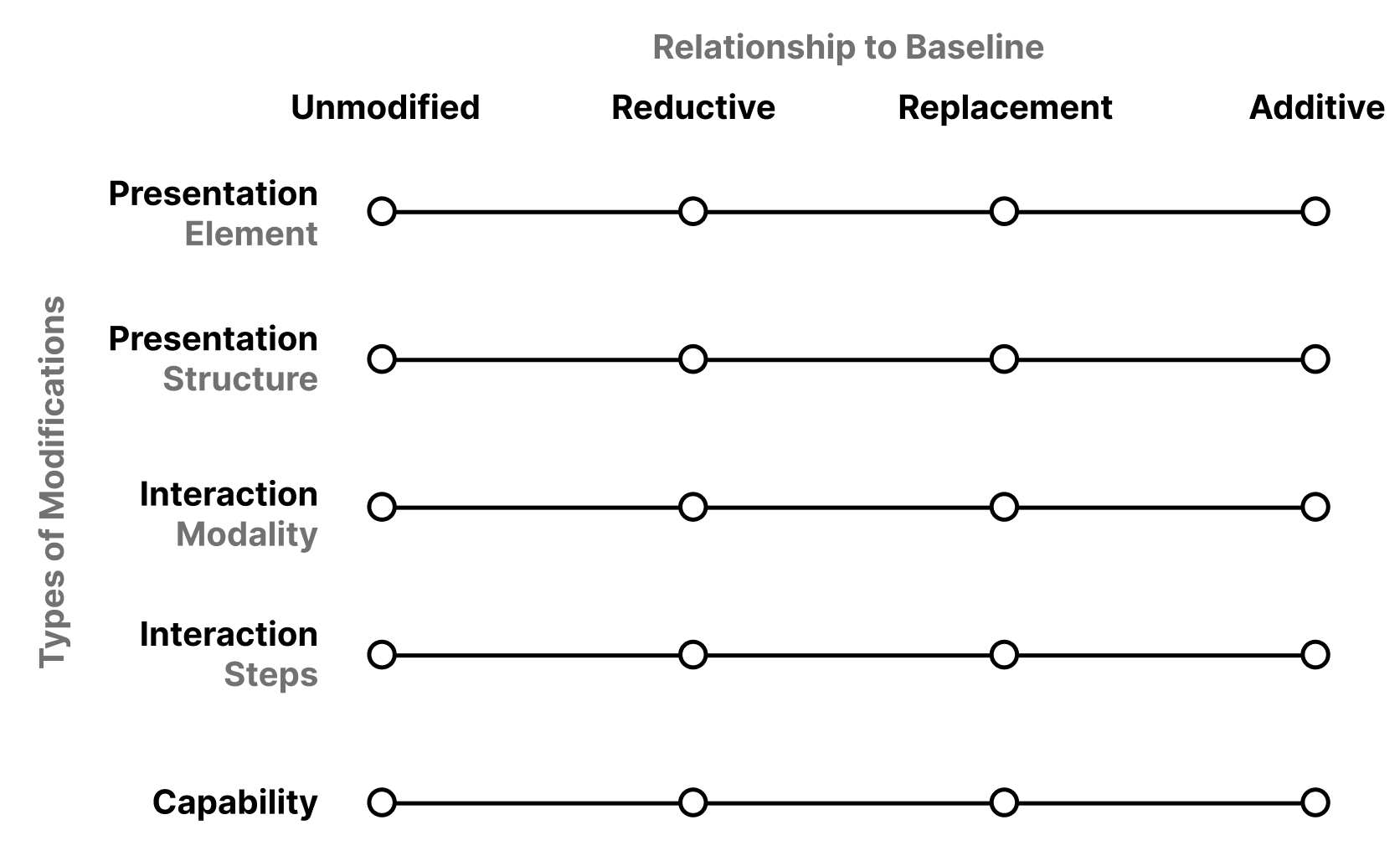}
    \caption{Design Space of Software Modification. \textbf{Types of Modifications} describes the area of user experience modified. \textbf{Relationship to Baseline} describes the way the modified experience relates to the unmodified software.}
    \label{fig:concept-what}
\end{figure}

The design space (\autoref{fig:concept-what}) comprises two axes. 

First, Types of Modifications describes what part of the user's experience with the target application has changed, from what users see (\textit{presentation}), how the user operates the application (\textit{interaction}), and what they can do with the application (\textit{capability}). A presentation change can affect individual elements, for example, labels and icons (\textit{element}) or the way elements relate to one another (\textit{structure}). Similarly, an interaction change can affect individual action (\textit{modality}) or a sequence of actions required to complete a task (\textit{steps}). Capability changes whether a functionality is available to the user at all. Our dimensions follow similar distinctions in prior design spaces that distinguish transformations of objects from their layouts~\cite{florinsGracefulDegradationUser2004}, transformation from presentation~\cite{minMakingAbstractionConcrete2026}, and behaviour from presentation~\cite{beaudouin-lafonDesigningInteractionNot2004}.

Second, Relationship to Baseline describes how a modification relates to the baseline for each dimension. First, the baseline can be \textit{unmodified}. A change can be \textit{reductive}, removing something from the baseline, \textit{additive}, introducing something new, or \textit{replacement}, where something is removed and swapped with something new. Our mappings follow the same descriptors found in prior design space of text manipulation~\cite{zindulkaExploringMobileTouch2025} and the design space of interaction re-mappings~\cite{zhangInteractionProxiesRuntime2017}. The following sections explain dimensions with example systems and compare the dimensions with prior design spaces.

\subsubsection{Presentation Element}

Concerns the visual properties of an individual interface element. 
\textbf{Reductive} removes part of an element's presentation without providing alternative encoding in its place. For example, reducing the saliency like \textit{ScreenCrayons} rendering the unannotated areas with a light colour~\cite{olsenScreenCrayonsAnnotatingAnything2004} and \textit{Ephemeral Adaptation} fading the menu items users are less likely to use~\cite{findlaterEphemeralAdaptationUse2009}. Adaptive hypermedia similarly describes elements being dimmed~\cite{debraAdaptiveHypermediaSystems1999} and Min et al.\ identify element representation that removes or simplifies encoding~\cite{minMakingAbstractionConcrete2026}. 
\textbf{Replacement} preserves the information carried by an element but substitutes another presentation. For example, \textit{Prefab} translates interface labels into another language~\cite{dixonContentHierarchyPixelbased2011}, \textit{Stylette} rewrites the cascading style sheets (CSS) of a selected web element from natural language~\cite{kimStyletteStylingWeb2022}, and \textit{DOMSteer} changes the border to increase the saliency of a web element~\cite{haoChatClicksGUI2026}.
\textbf{Additive} introduces new visual channels. For example, \textit{Prefab}'s favourites enhancement displays a star on existing widgets~\cite{dixonContentHierarchyPixelbased2011}, \textit{ToolClips} shows contextual tutorial video over AutoCAD icons~\cite{grossmanToolClipsInvestigationContextual2010}, and \textit{Sikuli} attach annotations to interface elements~\cite{yehSikuliUsingGUI2009}. The difference between replacement and additive is whether the original presentation remains. Both require identifying the elements, but the former also requires covering or redrawing the element. Presentation change is often also associated with capability change. For example, \textit{Patina} overlays community usage over the applications, providing new information while adding presentation element~\cite{matejkaPatinaDynamicHeatmaps2013} and \textit{reform} attaches UI enhancements with new information to a website~\cite{toomimAttachingUIEnhancements2009}. Both appear as presentation changes, but the information added did not previously exist in the original application, and we therefore consider them as capability additive as well.

\subsubsection{Presentation Structure}

Concerns the organization among the elements, such as the position, grouping, and distribution across regions or screens. 
\textbf{Reductive} hides groups of elements from the user. For example, the personal interface for Word changes the commands exposed by its menus and menu bars~\cite{mcgrenereEvaluationMultipleInterface2002}, \textit{Dark Pita} enables users to suppress distracting interface content~\cite{luAwarenessActionExploring2024}. In these cases, the individual hiding is also considered a reducing presentation element.
\textbf{Replacement} restructures elements. For example, \textit{Interface Facade} lets users cut, duplicate and recombine regions of widgets of running applications~\cite{stuerzlingerUserInterfaceFacades2006}, \textit{Masonview} reorganizes elements of existing websites~\cite{minDemonstrationMasonviewContentDriven2023}, and \textit{Orca}~\cite{jiangOrcaBrowsingScale2025} and \textit{Chickenfoot}~\cite{millerRewritingWebChickenfoot2010} rewrites the layout of web content. \textit{SUPPLE} restructures an interface by computing a layout based on device and user models~\cite{gajosSUPPLEAutomaticallyGenerating2004}. 
\textbf{Additive} appends a new region alongside the existing layout. For example, \textit{WinCuts} copies a live region of a window into another view without removing it from the source window~\cite{tanWinCutsManipulatingArbitrary2004}, \textit{TaskLens} add generated task-oriented panels beside Blender's existing interface for learning~\cite{liuTaskLensGeneratingTaskConditioned2026}, and Gajos et al.'s split interface design creates a duplicate additional toolbar while keeping the unmodified interface~\cite{gajosExploringDesignSpace2006}. Similar to presentation element modification, structural modification may also introduce new capabilities. For example, \textit{Fusion}'s mashed-up UI~\cite{zhangFusionOpportunisticWeb2018} and \textit{Wildcard}'s spreadsheet~\cite{littWildcardSpreadsheetdrivenCustomization2020} appear as additional panels but also introduce computation (capability) not previously available in the application. 

\subsubsection{Interaction Modality}

Concerns about modifying the type of modality for an interaction.
\textbf{Reductive} disables a type of interaction. For example, disabling short button presses for accessibility~\cite{levineMouseAdapterPeople2005}, disabling touch input when a pen is present~\cite{hinckleyPenTouchNew2010}, or disabling single-finger taps to curb phone use~\cite{luInteractOutLeveragingInteraction2024}. 
\textbf{Replacement} swaps a type of interaction with another operation or pathway. For example, \textit{Prefab} adds target-aware pointing techniques~\cite{dixonPrefabImplementingAdvanced2010} or changes clicks to sliding gestures for tablet input~\cite{dixonPixelbasedMethodsWidget2014}. 
\textbf{Additive} introduces an alias to activate the same functions with new pathways. For example, the \textit{ICON} toolkit enables rerouting alternative input device to control an application~\cite{dragicevicSupportInputAdaptability2004}, \textit{Grab~'n'~Drop} adds toolglass interaction over third-party applications without removing the original menus~\cite{eaganGrabDropUser2017}, \textit{Astral} maps phone sensors to mouse and keyboard events for rapid prototyping~\cite{ledoAstralPrototypingMobile2019}, \textit{Genie} exposes recovered web commands through speech and keyboard without removing the page's existing controls~\cite{swearnginGenieInputRetargeting2017}. \textit{Improv}~\cite{chenImprovInputFramework2017} describes a system for modelling UI behaviours for remapping controls, but since the original interaction channel is intact, we consider it as additive. Zhang et al.'s one-to-zero, one-to-one, and zero-to-one mappings are similar to our reductive, replacement, and additive modifications~\cite{zhangInteractionProxiesRuntime2017}.

\subsubsection{Interaction Steps}

Concerns the sequence of carrying out a task. This is a view on change to the number of steps, which is different from prior definitions that concerned adding or removing elements, structures, or pathways. We take on this definition to echo the many-to-one and one-to-many interaction mappings in \textit{Interaction Proxies}~\cite{zhangInteractionProxiesRuntime2017}, as it helps describe automation.
\textbf{Reductive} shortens the steps it takes to complete an action without changing the action's outcome. Automating the systems shortens the number of steps for users to complete a task. \textit{CoScripter}~\cite{leshedCoScripterAutomatingSharing2008}, \textit{Chickenfoot}~\cite{bolinAutomationCustomizationRendered2005}, \textit{Sikuli}~\cite{yehSikuliUsingGUI2009}, \textit{SUGILITE}~\cite{liSUGILITECreatingMultimodal2017}, \textit{Koala}~\cite{littleKoalaCaptureShare2007}, \textit{MultiClick}~\cite{zhangMultiClickCrossTabWeb2025}, \textit{PUMICE}~\cite{liPUMICEMultiModalAgent2019}, and \textit{Rousillon}~\cite{chasinsRousillonScrapingDistributed2018} all convert repetitive or multi-step interactions into fewer steps or a single trigger. When the system creates a new trigger, such as a new button, that bypasses the original interaction steps, it is also considered to have modified the interaction modality. \textit{Anchored Customization}~\cite{ponsardAnchoredCustomizationAnchoring2016} surfaces, anchors, and highlights setting options in a multi-layered interface, and in doing so, it also modifies presentation elements (adds a highlight) and structure (adds controls).
\textbf{Replacement} reroutes one sequence with another without necessarily shortening it. For example, \textit{Interface Facade}~\cite{stuerzlingerUserInterfaceFacades2006} reroutes a drop-down menu with radio buttons, but this does not change the number of steps to complete the operation.
\textbf{Additive} adds steps and sequences to the interaction. For example, \textit{InteractOut} adds requirements for repeated taps to trigger an action~\cite{luInteractOutLeveragingInteraction2024} and \textit{Interaction Proxies} replaces one interaction with a sequence of two interactions~\cite{zhangInteractionProxiesRuntime2017}, 

\subsubsection{Capability}

Concerns the set of outcomes an application makes available.
\textbf{Reductive} restricts the outcome of the application to what's reachable from the user. UI mashups extract capabilities from an existing application to package them into an app. For example, \textit{A-Mash} extracts the selected functionality from an existing application~\cite{leeAmashProvidingSingleapp2022}. In this case, the application being extracted experiences reductive modification, as some of its functionalities are rendered unreachable. But for the newly formed mashup interface, it experiences additive modification. \textit{Varv}'s schema and actions-based application enables users to remove functionalities through either using the ``omit'' operator that removes functions or the ``pick'' operator that selects a subset of actions~\cite{borowskiVarvReprogrammableInteractive2022}. Since the end result is a reduced set of available features, we consider both as a reductive modification to the baseline.
\textbf{Replacement} overrides the underlying implementation and algorithm, which makes the performance or existing capabilities better, without introducing a new capability. For example, \textit{Chrome Frame} is an Internet Explorer plug-in that changes the browser's rendering engine while retaining the original browser interface~\cite{ChromeFrame}.
\textbf{Additive} extends the possible outcome from using the application. For example, \textit{WADE} adds external service into a host program~\cite{mengWADESimplifiedGUI2014}, \textit{ToolTrack} add counters to buttons to support users exploring and learning features~\cite{massonSuperchargingTrialandErrorLearning2022}, and computation abstractions add new capabilities to existing data~\cite{minMalleableOverviewDetailInterfaces2025}. \textit{Scotty} injects functionality to the send button to check if the user forgot the attachment~\cite{eaganCrackingCocoaNut2011}. This is an example of adding new capabilities but maintaining the original interface.

\subsection{Framework of Modification Mechanism}
\label{sec:framework}

We characterize how prior modification systems use different forms of \textit{application access} provided by the target application to achieve various levels of \textit{synchronization}. By that, we mean the extent to which the modified experience remains synchronized with the target application's state and actions during use. Tight coupling is particularly important for modification systems that leave the application's underlying implementation unchanged. Because these systems modify how users perceive and interact with the application rather than rewriting the application's code, the modified interface must continuously create a proxy to reflect the state of the target application, and vice versa. The tighter this coupling, the more the modification can behave as an integrated part of the application, allowing users to interact with the modified experience without distinguishing between the modification interface and the target application.

\begin{figure*}
    \centering
    \includegraphics[width=0.7\linewidth]{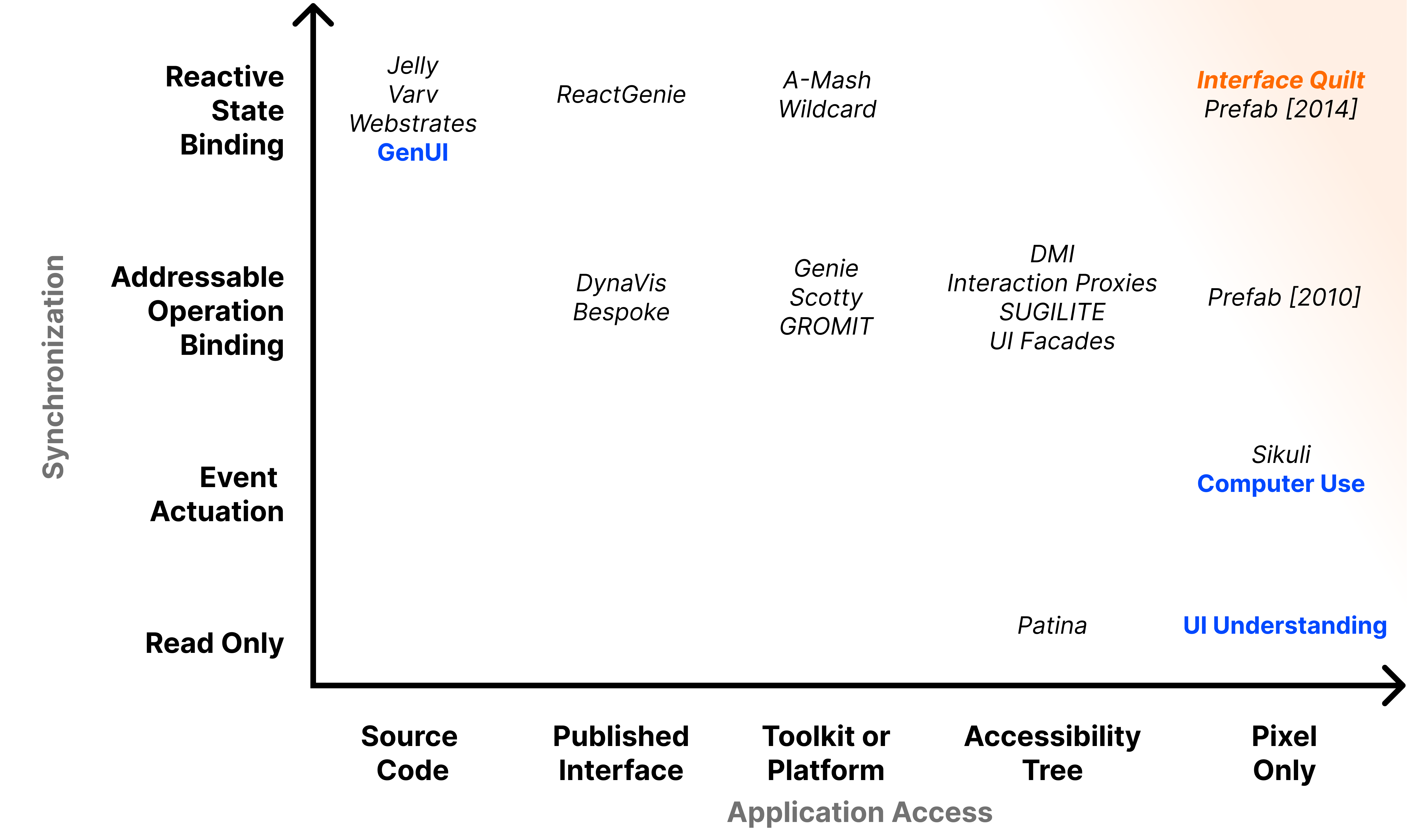}
    \caption{Framework of Modification Mechanism. \textbf{Application Access} describes the pathway used by the modification system. \textbf{Synchronization} describes the extent to which the modified experience remains synchronized with the target application's state and actions.}
    \label{fig:concept-how}
\end{figure*}

\subsubsection{Application access} 

This describes the representation of the target application that a modification system has access to.

\textbf{Source Code.} At one extreme, the modifying system owns the substrate to be modified. In other words, the modification system is the application. As such, the presentation, data, and representations can all be directly inspected and edited by the system. \textit{Webstrates}~\cite{klokmoseWebstratesShareableDynamic2015}, for example, makes the running application a shared document that itself can rewrite, while \textit{Varv} represents applications through schema and actions that can be extended or omitted~\cite{borowskiVarvReprogrammableInteractive2022}. Because the modifying system controls the representation that drives the interface, systems can change the presentation, interaction, and capabilities directly.

\textbf{Published Interface.} This includes the domain representation provided by the developer, such as APIs, MCPs, command-line interface, scripting, and plugin frameworks. For example, \textit{DynaVis} interfaces with Vega-Lite to generate widgets with callbacks to modify properties of the visualization~\cite{vaithilingamDynaVisDynamicallySynthesized2024}, and \textit{Bespoke} builds controls over command-line applications~\cite{vaithilingamBespokeInteractivelySynthesizing2019}. This provides a stable access to tested controls. However, the generalizability of modifications would be limited by what the developers enable.

\textbf{Toolkit or Platform.} A modifying system may choose to exploit the UI toolkit and or features of the programming language that were not designed as extension points. For example, \textit{Genie} analyze a host application's JavaScript and event structure~\cite{swearnginGenieInputRetargeting2017}, \textit{Scotty} changes method implementations in running Cocoa applications through runtime toolkit overloading~\cite{eaganCrackingCocoaNut2011}, and systems such as \textit{A-Mash} use operating-system mechanisms to extract and embed live interface components from unmodified applications~\cite{leeAmashProvidingSingleapp2022}. These accesses are not intentionally designed by the app developers, but can be used by the modification software if the application is built with those toolkits and languages.

\textbf{Accessibility Tree.} It exposes a structured description of an interface outside the application's implementation. Systems such as \textit{Interaction Proxies}~\cite{zhangInteractionProxiesRuntime2017} and \textit{Interface Facades}~\cite{stuerzlingerUserInterfaceFacades2006} use accessibility information representation to recover element labels, roles, bounding boxes, states, and support actuation. These are routes that provide some access to the application internals, but availability is dependent on the quality and completeness of the accessibility information supplied by application developers. 

\textbf{Pixel.} Finally, a modification system can use the same visible output channel that's available to the user. \textit{Prefab}~\cite{dixonPrefabLayersPrefab2014} locates the controls by matching pixel-based pixel properties and recovers widget state and style from the hand annotations~\cite{dixonPixelbasedMethodsWidget2014}, and \textit{Sikuli} similarly locates click targets from screenshot templates~\cite{yehSikuliUsingGUI2009}. While rendered output requires the least cooperation from the application developer, it demands the most effort for a modifying system to recover the structure, behaviour, and states that were explicitly provided.

Moving from the left (source code) to the right (pixel) poses a trade-off between developer-supported structure and broader applicability of modifications. Improvements in UI screen understanding and computer use models make it increasingly easy to leverage pixel as a modification substrate, but the access still needs to be connected with actuation back to the unmodified application.

\subsubsection{Synchronization}

Synchronization describes the extent to which the modified experience remains synchronized with the target application's state and actions at runtime. We distinguish levels of coupling by how directly and continuously this synchronization can be maintained. The mechanisms supporting that synchronization may be explicitly exposed by the target application, or reconstructed by the modifying system from less structured forms of access.

\textbf{Read Only.} This is the weakest level, where a system can only recover information about the current interface without routes to actuate on the host software. UI-understanding systems, for example, recover interface elements from pixels~\cite{zhangScreenRecognitionCreating2021,wangScreen2WordsAutomaticMobile2021,
wuScreenParsingReverse2021} and \textit{Patina} uses the accessibility API to recognize buttons to overlay dynamic usage visualization without actuating on the application~\cite{matejkaPatinaDynamicHeatmaps2013}. The reading channel can inform a presentation modification, but cannot affect the interaction.

\textbf{Event Actuation}. This adds another route through user-level input. For example, \textit{Sikuli} locates the targets from screenshots and automates executions through synthetic mouse and keyboard events~\cite{yehSikuliUsingGUI2009}. More recent vision-based computer use agents, such as \textit{UI-TARS}, operate applications via unit actions like clicks, typing, and scrolling~\cite{qinUITARSPioneeringAutomated2025}. This enables OS-level actions over an application that is opaque for modification. However, input events do not by themselves provide stable handles to operate the host application, and the system must continuously decide where and how to act.

\textbf{Addressable Operation Binding.} This provides the modification with a stable handle to the application. Compared with the previous level, it provides modifications that identify control, procedure, and steps, enabling the modifications to affect the operations of the software. The handle might be an interface control, such as \textit{Interaction Proxies}' accessibility nodes as inspectable state and route to invoke actions on an element~\cite{zhangInteractionProxiesRuntime2017}. Or the handle can be procedures and commands. \textit{SUGILITE} generalizes demonstrated interactions into reusable procedures~\cite{liSUGILITECreatingMultimodal2017} and the \textit{Declarative Model Interface} abstracts GUI interaction into explicit access, state, and observation primitives for agents to operate~\cite{wangImperativeDeclarativeLLMfriendly2026}.

\textbf{Reactive State Binding.} The strongest operational coupling is to be able to both manipulate the host software and synchronize response to changes to the modified interface to promote human use. For example, \textit{Wildcard}'s spreadsheet is synchronized with an application-editable spreadsheet, and changes in one representation propagate between representations~\cite{littWildcardSpreadsheetdrivenCustomization2020}. When the modifying system owns an underlying representation, this type of coupling is provided for free. For example, \textit{Jelly} can modify a shared task-driven data model to drive real-time updates to the data and generate an interface~\cite{caoGenerativeMalleableUser2025}. Reactive binding is also possible with a pixel alone. The \textit{Prefab} follow-up paper synchronizes the button states between the original interface and a sliding widgets overlay through pixel-based recognition~\cite{dixonPrefabLayersPrefab2014}.

\section{Interface Quilt and Interface Patch}
\label{sec:demos}

We propose Interface Quilt and interface patches for modifying existing applications (\autoref{fig:teaser}). An interface patch is a generated interface overlaid on the existing application to modify the user’s experience with the application's presentation, interaction, and capabilities. And Interface Quilt is a system that supports generating and running interface patches through computer use agents (CUAs) and generative UI (GenUI).

\subsection{Interface Quilt Supports Generating and Running Interface Patches}

The design space (\autoref{fig:concept-what}) identifies modifications to an application's presentation, interaction, and capabilities, and the mechanism framework (\autoref{fig:concept-how}) characterizes how such modifications access and maintain synchronization with the application. These analyses informed the design of \sys{}.

\sys{} consists of two parts. First, a build-time pipeline uses CUAs to explore the target application and identify the available UI controls and observable states required to access and synchronize an interface patch with the application. A coding agent then uses these identified operations to generate an interface patch. Second, a runtime hosts the generated interface patch alongside the target application. When a user interacts with the interface patch, the runtime executes the corresponding UI actions in the target application. When relevant application state changes, \sys{} observes those changes and updates the interface patch accordingly.

This architecture reflects two points in our framework (\autoref{fig:concept-how}). First, \sys{} recovers the target application through its rendered interface rather than source code, application APIs, or accessibility metadata. Second, it maintains synchronization between the target application and the interface patch. Interface patches can request application operations, and application state can update the patch. Together, these mechanisms allow the same runtime to support interface patches that modify presentation, interaction, and capability, which we demonstrate in the following sections.

\subsection{Demonstrating Interface Patch Modifying Applications}

We first provide two compound examples, \`{A} la Carte Interface and Put-that-there, to demonstrate the interface patch's ability to execute multiple types of modifications at once. We then provide two sets of interface patch examples targeting multiple applications to showcase specific types of modifications.

\subsubsection{\`{A} la Carte Interface}

\begin{figure*}
    \centering
    \includegraphics[width=1\linewidth]{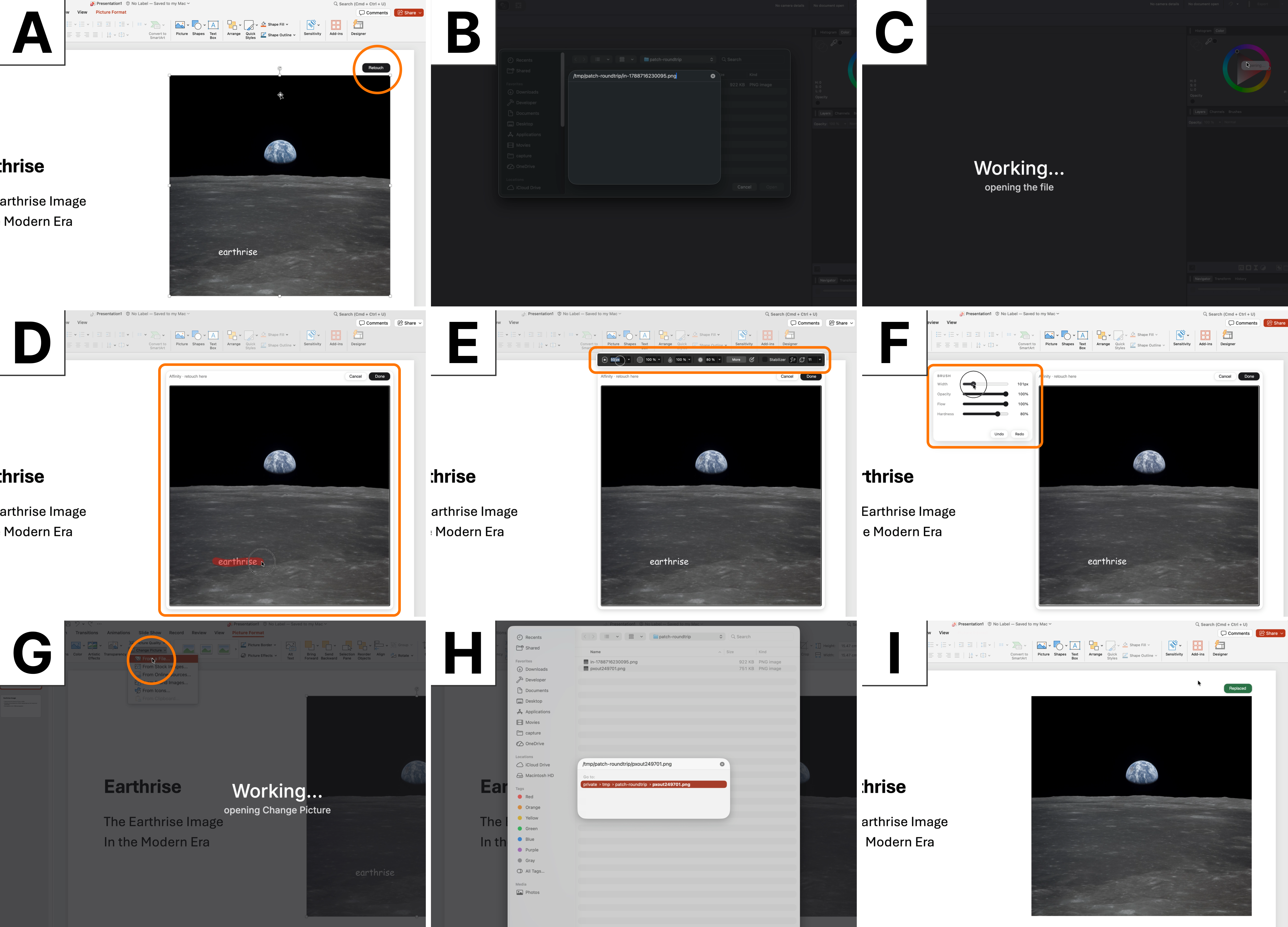}
    \caption{Interface Patch Example: Revising WinCut and Interface Facade with an \`{A} la Carte Interface. Supporting editing images in PowerPoint with the features of Affinity. (A) User clicks the retouch button. (B) System opens a file from a path. (C) System clicking to open Affinity. (D) User retouches the image on the interface patch cutout. (E) Cutout of the original interface's control. (F) Generative UI replaces the original interface's control. (G) The system executes the pre-discovered execution sequence to operate the PowerPoint and replace files. (H) System opens a file from a path. (I) The retouched image is replaced in PowerPoint. Orange outline highlights the interface patch's interface and actions and is not part of the user's experience.}
    \label{fig:demo-roundtrip}
\end{figure*}

\autoref{fig:demo-roundtrip} shows an interface patch surfacing only the relevant interface and controls to the user. The user is making a slide deck in PowerPoint and wants to retouch an image to remove the text. With this interface patch, they can tap into the retouching algorithm of professional software like Affinity, without needing to leave PowerPoint, manage multiple open windows, sync the image between applications, or learn how to use the complex interface of Affinity. The user interacts with an interface patch cutout of Affinity's canvas and retouches by gesturing to paint like they usually do (\autoref{fig:demo-roundtrip}D-F). \sys{} supports generating UIs, enabling different representations of controls and gradually revealing relevant controls for the user's task, ranging from no settings (\autoref{fig:demo-roundtrip}D), a cutout of original tool settings (\autoref{fig:demo-roundtrip}E), to a reksinned set of settings with sliders that synchronize with the target application (\autoref{fig:demo-roundtrip}F).

This workflow demonstrates \sys{}'s ability to create an interface patch that modifies across the design space. On presentation, new interface elements are additive, and the cutout of the original Affinity interface is a structural addition. On interaction, the user's paint operation on the interface patch is unmodified, the GenUI sliders are modality replaced, and the synchronization between the PowerPoint image and Affinity is step reduction. On capability: from Affinity's perspective, this is a capability reduction because the user can no longer access other features of Affinity through this interface patch, whereas from PowerPoint's perspective, this is both a capability replacement, using Affinity's algorithm for retouch, and a capability add, as PowerPoint's erase feature does not contain settings for the retouch pen.

This is an extended version of prior systems like WinCuts~\cite{tanWinCutsManipulatingArbitrary2004}, Interface Facade~\cite{stuerzlingerUserInterfaceFacades2006}, and Highlight~\cite{nicholsHighlightSystemCreating2008}. In addition to duplicating interface elements and screen regions, \sys{} enables (1) packaging interaction paths discovered from computer use agents at build time into a single button-triggered deterministic workflow at runtime. This effectively breaks down the cross-application data silos; (2) leveraging the generative UI capabilities to reskin one applications' UI for new ways to interact and more cohesive integration to another application; and (3) automatically handling the window management to position the interface patch directly above the original image in PowerPoint, creating an integrated experience.

\subsubsection{Put-that-there}

\begin{figure*}
    \centering
    \includegraphics[width=1\linewidth]{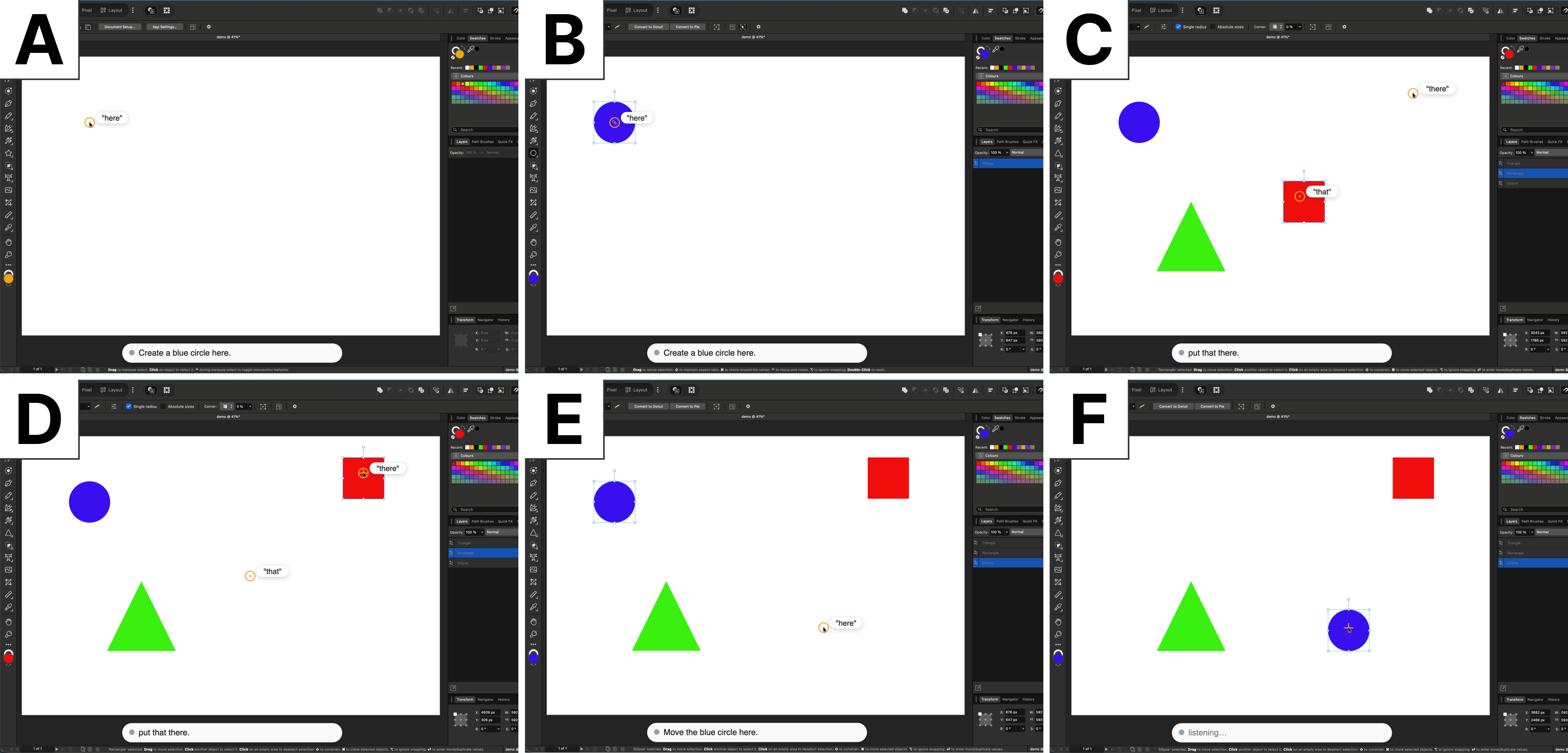}
    \caption{Interface patch example: Put-that-there. (A) User says ``create a blue circle here'' while pointing the cursor at a point on the canvas. (B) The system creates a blue circle. (C) User says ``put that'' while pointing at the red square and ``there'' while pointing at another point on the canvas. (D) The red square moves to where the user pointed. (E) User says ``move the blue circle here'' while pointing on the canvas. (F) The system identifies the object and moves it to where the user points.}
    \label{fig:demo-ptt}
\end{figure*}

\autoref{fig:demo-ptt} shows the user manipulating shapes on the vector graphics app, Affinity, with their voice and cursor position, inspired by the put-that-there interaction technique~\cite{boltPutthatthereVoiceGesture1980}. With \sys{}, the user can operate Affinity through multimodal interactions even when the software does not support it. The interface patch includes a live speech bubble at the bottom of the interface. The interface patch also renders the user's cursor location and corresponds to speech on the canvas, like ``here'' and ``there.'' 

To create an object, the user describes its properties while pointing to the desired location. The interface patch combines the speech and pointing input and maps the request to application operations recovered at build time. \sys{} then invokes the corresponding Affinity controls to create the shape (\autoref{fig:demo-ptt}A–B). To move an object, the user says ``put that'' while pointing to the object and ``there'' while pointing to its destination. \sys{} recovers the referenced object's location, while the interface patch interprets the user's multimodal command and invokes the corresponding move operation. The user can refer to the object by location or semantics. For example, the user can say ``move the blue circle here.'' The interface patch identifies the object and its coordinates, then moves the object to the user's pointed location. In addition, by leveraging \sys{}'s ability to identify object location, this put-that-there interface patch also supports adding different shapes with different sizes, changing the colour of a shape, multi-select, align, delete and duplicate shapes.

This demonstrates \sys{}'s ability to create an interface patch with a new presentation element. As the user can still operate the original interface as is, the interface patch does not replace the original interaction but adds an alternative voice input modality. It reduces interaction steps, because it automates procedures such as adding objects and changing colours into one command. As the capabilities for rendering and manipulating objects remain the same, this interface patch does not change the application's capabilities.

Compared with the original put-that-there~\cite{boltPutthatthereVoiceGesture1980} implementation for large screen multimodal interactions, our system uses the same point and speech interactions for a desktop application. Because the interface patch is implemented by supplying the coding agent with the original paper, it uses the same symbolic matching algorithm for identifying objects. For recognition of shapes and transcribing the user's speech, the interface patch sends requests via external APIs. The experience of using the interface patch compared with a vanilla computer use agent is that the interface patch retains the mix-initiative experience, where the user provides verbal commands while maintaining precise location input via cursor location, and repetitive steps like selecting, adding, and aligning multiple objects are automated.

\subsubsection{Many Interface Patches, Single App}

\begin{figure*}
    \centering
    \includegraphics[width=1\linewidth]{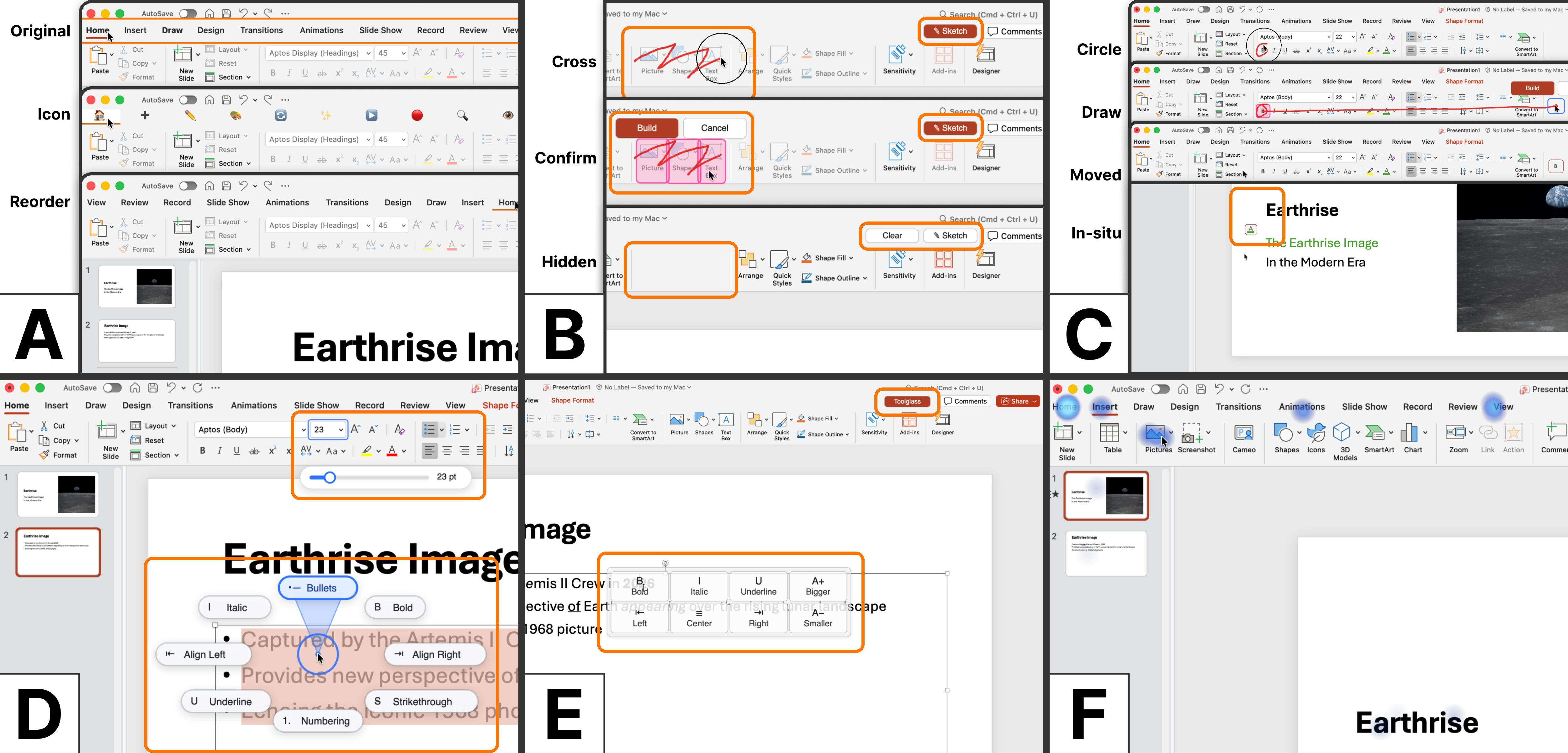}
    \caption{Interface patch example: Many Interface Patches targeting PowerPoint. (A) Change ribbon to icon or reorder icons. (B) Inspired by SILK, sketch to remove the element. (C) Sketch to duplicate buttons. (D) Text size slider and marking menus. (E) Toolglass technique. (F) Patina technique. Orange outline highlights the interface patch's interface and actions and is not part of the user's experience.}
    \label{fig:demo-ppt}
\end{figure*}

\autoref{fig:demo-ppt} shows many interface patches modifying an application's presentation, interaction, and capabilities. 

\textbf{Iconify and Reorder}. Sometimes interface presentation or layouts are not optimal for the user's task, and \autoref{fig:demo-ppt}A shows two interface patches changing the presentation element and structure. The first changes the original PowerPoint ribbon text buttons into icons. The second reorders the ribbon items while preserving the original visual language. At build time, \sys{} recovers controls' locations and visual properties and reassembles them as an interface patch. At runtime, the \sys{} accepts the user's interaction with the interface patch and operates the original interface.

\textbf{Sketching Interface}. \sys{} can also include an authoring interface patch that allows users to modify an application's presentation. Inspired by Landay's \textit{SILK}~\cite{landaySILKSketchingInterfaces1996} that enables users to quickly sketch out interactive interface prototypes, \autoref{fig:demo-ppt}B and C show an interface patch that enables users to hide unwanted interface elements and duplicate buttons, respectively. To hide, the user first presses the ``Sketch'' button that shares the same visual language as PowerPoint, then zig-zags with the cursor over the unwanted elements. By pressing ``Build,'' the interface patch applies a colour mask to hide the elements (\autoref{fig:demo-ppt}B). The user can also move buttons around by circling an element, draw where the element should go, and have the button duplicated to the new area on the ribbon or on the canvas for in-situ controls (\autoref{fig:demo-ppt}C). This interface patch demonstrates presentation structure reduction (hide buttons) and addition (duplicate buttons), and capability addition (user authoring element changes at runtime).

\textbf{Slider and Marking Menus}. HCI has invented many advanced interaction techniques that are not implemented in all software. \sys{} enables recreating and attaching these techniques to existing software via interface patches. Inspired by Kurtenbach and Buxton~\cite{kurtenbachLimitsExpertPerformance1993}, \autoref{fig:demo-ppt}D shows the interface patch's Marking Menus, which allow users to right-click on the canvas to open the radial buttons representing the menus. The user can perform gestures towards an icon to apply the menu option, or directly perform the gesture towards a desired control in one gesture, without pausing for the radial button to display. This allows the user to quickly apply styling options on the target application, such as performing a gesture to align text to the right. This is done by recovering the controls in the ribbon and pressing for the user as triggered by the marking menu interface patch. This example also shows \sys{}'s support for hosting multiple patches at once. For example, the slider interface patch provides an additional slider for adjusting text size. It continuously tracks the values in the text field, updates the values on the slider, and updates the user's input on the slider back into the text field, achieving the effect as if the controls are native. Both interface patches are modality additions.

\textbf{Toolglass}. Another classical interaction rarely implemented is \textit{Toolglass and Magic Lenses} by Bier et al.~\cite{bierToolglassMagicLenses1993}. \autoref{fig:demo-ppt}E shows an interface patch adding a toolglass to PowerPoint. After pressing the ``Toolglass'' button, the user can move the toolglass panel with the laptop's built-in trackpad and apply options with a mouse. This interface patch adds interaction modality.

\textbf{Patina}. Because Interface Quilt tracks the application state and the location of interface elements, it can associate users' interactions with the parts of the interface on which they occur. This enables interface patches that use interaction history to support learning the software. For example, inspired by Matejka et al.'s Patina~\cite{matejkaPatinaDynamicHeatmaps2013}, which visualizes usage patterns through heatmap overlays, \autoref{fig:demo-ppt}F visualizes the user's clicks on both UI elements and the canvas as a heatmap. When the user switches between tabs in the ribbon, the patch displays the corresponding patina for the currently visible interface. Interface Quilt monitors the application through vision and maintains awareness of its changing state, allowing the visualization to remain bound to the relevant interface elements as the user interacts with the application. The interface patch adds a presentation element and capability to the application.

\subsubsection{More Interface Patches, More Apps}

\begin{figure*}
    \centering
    \includegraphics[width=1\linewidth]{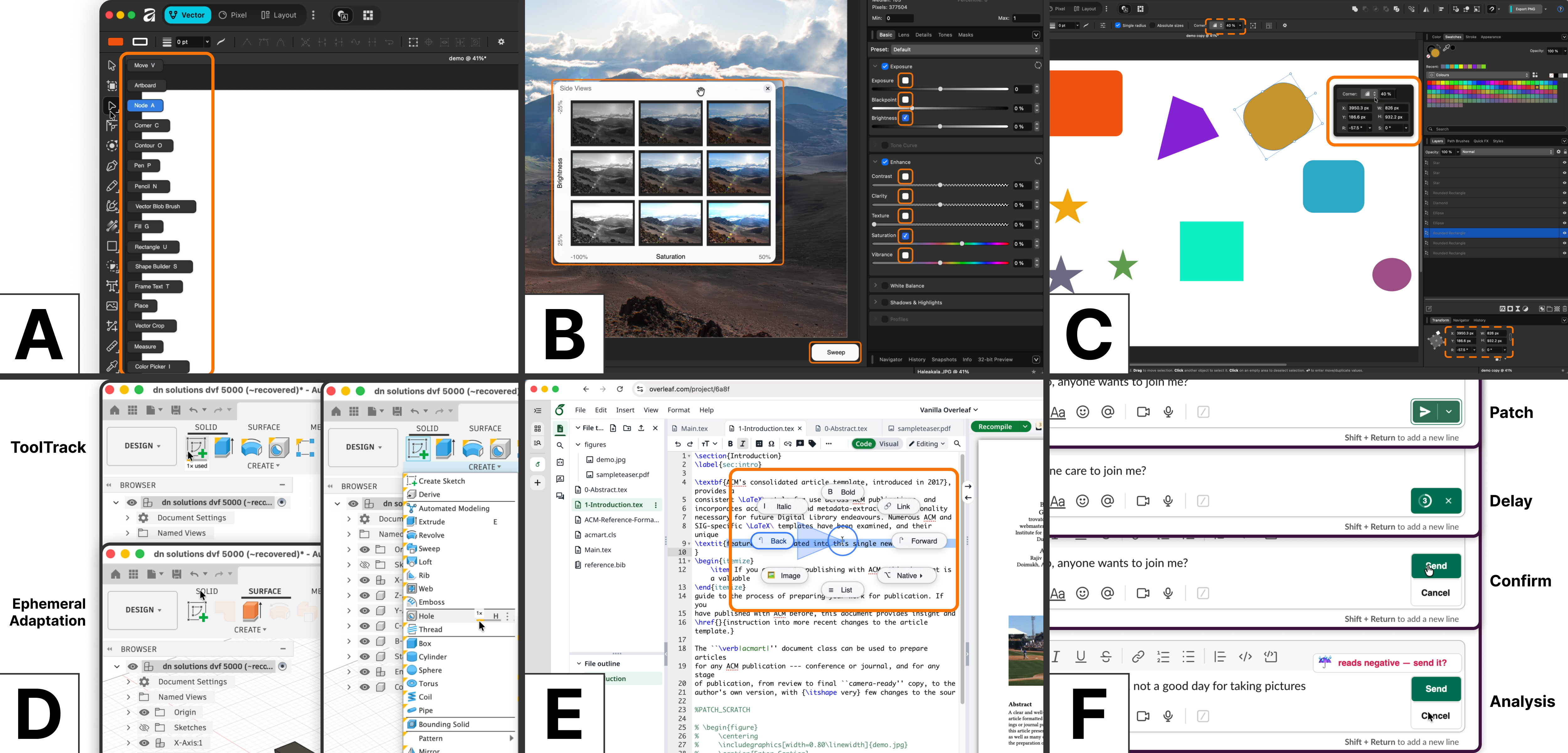}
    \caption{More Interface Patches. (A) Affinity Toolbar Labels. (B) Affinity 2D Side Views. (C) Affinity in-situ controls are composed of control panels from different regions of the interface. (D) Fusion ToolTrack and Ephemeral Adaptation. (E) Overleaf Marking Menu. (F) Slack enhancements on the send button, with delay send, confirm send, and sentiment analysis capabilities. Orange outline highlights the interface patch's interface and actions and is not part of the user's experience.}
    \label{fig:demo-more}
\end{figure*}

\autoref{fig:demo-more} shows more interface patches modifying more applications' presentation, interaction, and capabilities.

\textbf{Toolbar Labels}. Interfaces of complex software can be daunting and difficult to understand. Even though Affinity includes a tooltips implementation, where hovering over the icon shows the name of the tool, it remains hard for beginners to learn the icon-feature mappings. \autoref{fig:demo-more}A shows an interface patch that adds the name of the tool next to the icons and highlights the current tool in a more salient blue colour. \sys{} at build time, hover over each of the icons and record the tool names, and at runtime, it displays the name and synchronizes the selected tool as highlights. Since the original software contains the information but does not reveal it, this is a presentation element and structure addition, but not a capability addition.

\textbf{Side Views}. Another example of presentation added through feeding forward~\cite{vermeulenCrossingBridgeNormans2013} information contained in the software, but not immediately accessible by the user, is \textit{Side Views} by Terry and Mynatt~\cite{terrySideViewsPersistent2002}. \autoref{fig:demo-more}B shows an interface patch that allows the user to select preview dimensions from the interface patch's added checkbox, preview the possible outcomes of adjusting both dimensions, then apply the settings through the interface patch. Similar to Prefab's implementation of side views~\cite{dixonPrefabImplementingAdvanced2010}, \sys{} also sweeps the possible options by clicking on the slider. However, ours creates mechanisms for users to select dimensions and compose 2D side views.

\textbf{In-situ Menus}. When using complex software, the controls are often embedded in inspectors positioned on the edge of the interface. \autoref{fig:demo-more} shows \sys{} recognizing the selected shapes, then updates the interface patch to display a situated control next to the user's selected object composed of the relevant parameter fields. The interface patch is dynamic to the object selected. For example, when the user selects a shape without the corner radius option, the interface patch no longer displays it. This interface patch demonstrates a presentation structure that is context-aware of the user's selected object's type and the location of the object.

\textbf{ToolTrack and Ephemeral Adaptation}. \autoref{fig:demo-more}D shows an interface patch reimplementing Masson et al.'s ToolTrack feature on \textit{Fusion}~\cite{massonSuperchargingTrialandErrorLearning2022}. It adds visualizations, such as yellow triangles and progress bars, to help users track whether a feature has been explored. The original implementation was built into the application and had source code access. Here, \sys{} recognizes the buttons, including those in the drop-down, and adds user clicks tracking from the outside. Similarly, inspired by Findlater et al.'s \textit{Ephemeral Adaptation}~\cite{findlaterEphemeralAdaptationUse2009}, the interface patch tracks the user's most recently used buttons, showing those buttons immediately, while gradually fading in the others. This interface patch changes the presentation element to reduce saliency and is a capability addition.

\textbf{Marking Menus}. We can also apply the same interface patch built for one application to another application. \autoref{fig:demo-more} shows the marking menu interface patch operating on Overleaf on Chrome. Because \sys{} operates on rendered pixels, it views Overleaf like any other GUI application. At build time, \sys{} identifies commands such as undo, redo, and bold. At runtime, the \sys{} invokes these recovered commands when the user selects the corresponding menu item.

\textbf{Button Enhancements}. \autoref{fig:demo-more}F shows three interface patches, adding capabilities. \sys{} first recovers the function and visual appearance of Slack's send button, then creates interface patches with the same button appearance overlaid in Slack. This enables the interface patches to effectively override Slack's button interaction and capabilities without access to its source code. First, a delay send feature that counts down after the user presses the send button, and triggers in the background after a set duration. Second, a confirm send feature that adds interaction steps, requiring a secondary user confirmation. Third, a sentiment analysis feature that extends the confirm send to (1) parse and analyze the user's drafted message and (2) run sentiment analysis to help users double-check the tone before sending a message.

\section{Implementation}
\label{sec:system}

\begin{figure}
    \centering
    \includegraphics[width=1\linewidth]{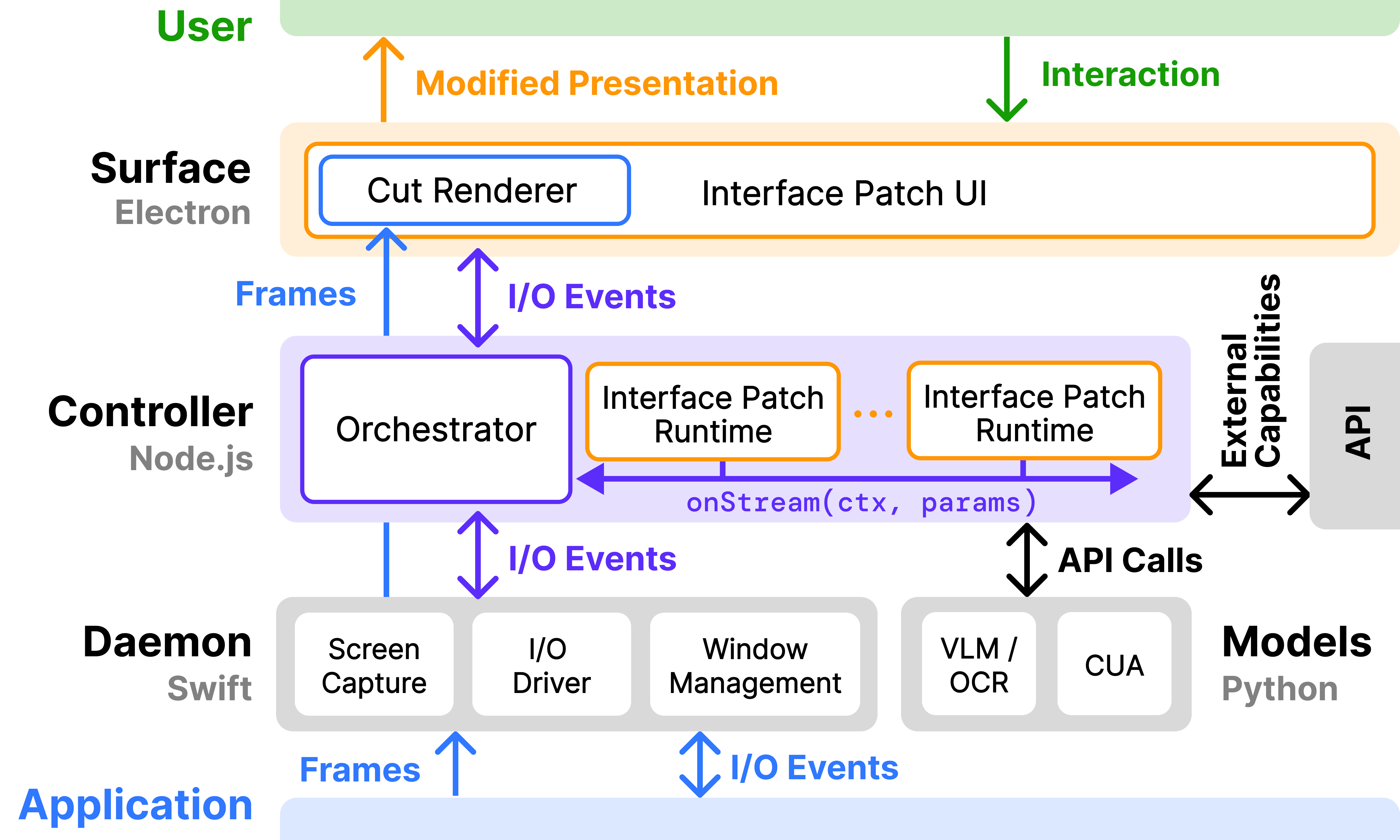}
    \caption{\sys{} Architecture.}
    \label{fig:arch}
\end{figure}

\sys{}'s runtime architecture separates the UI, logic, and OS support (\autoref{fig:arch}). The \textit{Surface} renders the interface patch UI and accepts the user's input. The \textit{Controller} orchestrates the I/O events, hosts multiple interface patch runtimes, and dispatches I/O events to the corresponding interface patches. The \textit{Daemon} and \textit{Local Models} are separate processes that provide an I/O connection for controllers with the OS and provide API call services to VLM/OCR and CUA.

We implement \sys{} on macOS, where the surface is an Electron window, the interface patches and controllers are written in TypeScript and executed in a Node.js runtime, and the daemon is written in Swift. We describe the implementation for macOS, but the same architecture can be applied to other operating systems.

\subsection{Interface Quilt Architecture}

\subsubsection{Surface Layer}
\label{sec:arch-surface}

The surface is a single Electron window displayed above all applications. We use Electron because it can reliably render the interactive interface patches written in TypeScript. By default, the surface is a transparent, always-on layer that permits all users' interactions to pass through to the application beneath it by setting \texttt{NSModalPanelWindowLevel}. Only the region where an interface patch is running becomes interactable. Having interface patches on a single surface that's controlled by the system enables precise interface element placement. We create a shared \textit{Cut Renderer} that contains shared frame streaming capabilities to avoid reimplementations between interface patches.

\subsubsection{Controller Layer}
\label{sec:arch-controller}

The controller is a standalone TypeScript Node.js process containing the runtime logic. It sits between the daemon and the surface, wiring OS-specific privileged operations such as capture and synthetic input via the daemon, while syncing the windows and hit regions with the surface. The controller decides on which route to best support an interface patch.

An \textit{orchestrator} manages all active interface patches by providing a central event queue \texttt{runStreams()} that provides each active interface patch with its own runtime, including event callbacks, drawing operations, hit-region namespace, and I/O requests. In the case of the slider and marking menu, the two interface patches each contain a runtime, but share the controller, surface, and daemon, managed by the orchestrator. Events from both the surface and the daemon return to the orchestrator. Native pointer, keyboard, window, and application change events come from the daemon, while interactions with the interface patch's UI come from the surface. The controller routes the events to the interface patch runtime that subscribes to it. For time-based events, where the interface patch needs to periodically poll the values on a region, such as syncing slider values, an interface patch can also initiate the poll without waiting for callbacks. In sum, the controller coordinates and schedules events rather than using an update loop, as many interface patches can be static and do not require periodic repaint.

\subsubsection{Daemon Layer and Local Models}
\label{sec:arch-daemon}

The daemon on macOS is a Swift application, packaging OS privileges for the controller access, containing the screen capture, an I/O driver, and window management modules. The decoupled design enables \sys{} to explore the OS interfaces once, and expose them as custom I/O events to service multiple interface patches. This avoids rebuilding or re-exploring the OS capabilities and enables sharing interface patches between OS by updating the daemon layer alone. 

The \textit{Screen Capture} module wraps macOS' \texttt{ScreenCaptureKit} for two purposes. First, to expose the UI for local models to recognize UI elements and parse text. Second, to support cutting and streaming UIs to interface patches without native support, like available on other OS~\cite{stuerzlingerUserInterfaceFacades2006}. The I/O driver handles clicking, typing, intercepting and notifying the orchestrator of low-level HID events.

The \textit{I/O Driver} posts events to the corresponding process either through \texttt{CGEvent.postToPid} or a private macOS API named \texttt{SLEventPostToPid}~\cite{cuaaiinc.Cua2025} to control the target application. This allows an interface patch to operate on a background application without moving the user's cursor or bringing the application to the forefront. However, the activation of menus and controls of some apps depends on the system pointer hover that cannot be simulated this way, such as PowerPoint's dropdown menus. In such cases, the daemon provides a fallback that temporarily borrows the user's cursor with \texttt{CGWarpMouseCursorPosition} to click. The daemon streams continuous user mouse events live to support the painting strokes. Synthetic events are tagged so that the daemon does not report them back to the controller. To type, the daemon uses \texttt{CGEvent(KeyboardEventSource:)} to simulate low-level keyboard events. 

The \textit{Window Management} module supports identifying which window is currently active with \texttt{CGWindowListCopyWindowInfo} to switch between corresponding interface patches. It also identifies the pid of the current window, allowing the display of the corresponding interface patch and the wiring of I/O events.

The \textit{Local Models} provide three paths for parsing the application's UI. First, a full pass uses bounding boxes, OCR, and computer vision models to identify UI elements and controls. Second, a faster, cheaper runtime route uses the UI grounding cached at build time to conduct pixel-based comparisons using local luminance for text tags and zero-mean normalized cross correlation (ZNCC) glyph template matching. Third, the daemon also uses OS-specific user input APIs that do not assume developer support as heuristics to trigger screen captures and state updates

\subsection{Building an Interface Patch}

\sys{}'s architecture serves as a development harness for coding agents. At build time, the coding agent uses \sys{}'s computer use and perception capabilities to explore the target application. We prompt a coding agent with the desired modification and give it access to the existing \sys{} codebase and runtime architecture. When recreating an interaction technique from prior work, such as marking menus or put-that-there, we additionally provide the corresponding paper as a specification of the intended behaviour. The coding agent can use \sys{}'s daemon to control the target application and inspect the streamed interface, while local models expose information such as element bounding boxes, text, and visual assets. It can also use its own visual understanding to interpret interface structure and behaviour. Together, these infrastructures allow the coding agent to explore the target application, identify relevant states and controls, and implement the runtime interface patch specifying the behaviours.

Each interface patch contains a TypeScript file with its executable interaction logic and a JSON binding containing application-specific data. The TypeScript implementation exposes an \texttt{onStream(ctx, params)} function that connects the patch to controller events at runtime, while the binding stores application-specific information such as visual assets and execution routes. Separating patch logic from application-specific bindings allows the same interface patch implementation to target multiple applications with minimal changes. For example, we use the same marking-menu logic for both PowerPoint and Overleaf. Through \texttt{onStream}, an interface patch receives a shared context and subscribes to callbacks such as \texttt{onPointer}, \texttt{onKey}, and \texttt{onReload}. The context also exposes surface and daemon APIs, allowing the interface patch to invoke operations such as 
\texttt{paint()}, \texttt{skyClickAt()}, \texttt{keyToPid()}, and \texttt{pointerAt()}.

\subsection{Apparatus and Models}
\label{sec:arch-models}

We built and tested \sys{} on a MacBook Pro with an M2 Pro processor running macOS 26.5.2. We used a combination of models in our setup. During build time, we use \textit{OmniParser v2}~\cite{luOmniParserPureVision2024} to find candidate control regions, Apple Vision OCR API to detect text labels and values, and an image-to-coordinates computer use model \texttt{mlx-community/Holo2-8B-bf16}~\cite{Holo2HcompanyCollection2026} for identifying buttons and controls' coordinates. We built the patches and the \sys{} with Claude Code harness, with \textit{Claude Opus 4.8, 5}, and \textit{Claude Fable 5, 5.1} models. We leveraged the coding agent's ability to inspect screenshots to name buttons without texts, test visuals of built patches, and inspect failures pointed out by the interface patch author. During runtime, we use the same set of models without the Claude Code models. A number of interface patches require access to live LLM recognition. The Slack sentiment analysis OCRs the current draft and uses \texttt{gpt-5.4-mini} for analysis. Put-that-there recognizes speech via \texttt{gpt-realtime-1.5} with \texttt{gpt-4o-mini-transcribe} for live transcription and uses \texttt{whisper-1} for timestamps matching the keywords and cursor positions.

\section{Discussion}
\label{sec:discussion}

We review the successes and limitations of \sys{} as a system for creating and running interface patches on applications with only pixel access, consider when alternative forms of application access could better support their construction and runtime synchronization, and discuss the broader implications of \sys{}'s architecture and interface patches as a form of software modification.

\subsection{What Interface Patches Can and Cannot Modify}

Our demonstrations show that interface patches can modify presentation, interaction, and capabilities across a range of existing applications, but that these modifications vary substantially in difficulty. These differences stem from what \sys{} can observe and control through the rendered application UI, and from how much of the target application's behaviour an interface patch must reproduce rather than reuse.

Adding \textit{presentation} and \textit{capabilities} is simple, as \sys{} can render a new interface running on top of the target application. It is harder to reduce or replace \textit{presentation}, because the system needs to first identify the area of the original interface through a vision model, then recover assets such as text and labels to recreate an interface patch with coherent visual language. 

Reducing and replacing the \textit{capabilities} is even more effortful. It is often easier to reuse the main canvas interface and recreate the controls surrounding it. For example, with \sys{}, it is possible to achieve the effect of a complete replacement of application A's core capabilities with a similar application B while retaining A's interface. However, to do this, \sys{} would recreate application A's interface and use it as controls for application B, rather than actually replacing the runtime library of application B. This is because \sys{}'s approach is suitable for recreating the controls UI as they are observable, but impossible to recreate an application's internal algorithm, such as rendering behaviours, as it's inaccessible from observation alone.

The difficulty of \textit{interaction} modifications varies by modalities or what is exposed on the surface of the target application. Mapping users' clicks or more sophisticated input techniques to basic clicks on the interface is easy, as the buttons are exposed and \sys{}'s daemon can easily reproduce clicks. Similarly, adding multimodal interactions is easy, as \sys{}'s architecture enables generating and hosting multimodal behaviours. Supporting gesture inputs on the target application is more difficult, such as a drag gesture, depending on its implementation. For example, we demonstrate success with Affinity (\autoref{fig:demo-roundtrip}), but we cannot retarget gestures on Fusion that used the Qt framework on macOS. The daemon has no path through the OS-level APIs to manipulate and stream the user's input events to Fusion, without taking over the user's cursor, which interrupts the user's sense of control.

\subsection{Pixels vs. APIs}

Pixels and computer use provide benefits not available through other application access pathways, such as API. First, the pixel provides \sys{} with visual information of the interface, enabling generating interfaces that share the visual language of the target application. We demonstrate this with the reordered controls (\autoref{fig:demo-ppt}A) and generated buttons with similar visual language positioned adjacent to the target application's controls (\autoref{fig:demo-ppt}B). Controlling the interface through computer use agents makes it easy to build multimodal interactions that require information about the user's cursor, like capturing the user's cursor in the put-that-there example (\autoref{fig:demo-ptt}).

The \sys{}'s pathway trades off seamless synchronization for generalizability. For example, in the \`{a} la carte interface patch, synchronizing the image between PowerPoint and Affinity required a sequence of clicks on the interface, rendering the OS temporarily unusable to the user. On one hand, APIs, MCPs, and scripting can make the process faster and work in the background. On the other hand, \sys{} promotes generalizability, where any application's file import and export can be executed, regardless of developer support. Additionally, even though the synchronization between the values on the interface patch and the target application is near instant, there exists a short delay. This becomes especially apparent in the in-situ control interface patch (\autoref{fig:demo-more}C) where \sys{} uses models to classify the object a user interacts with. We expect that more capable and efficient models will reduce or close the gap.

The difficulty of synchronizing the interface patch with the target application depends on the extent to which controls are exposed on the interface. It is easy when they are clearly marked. For example, the system can directly parse text and paint size values on PowerPoint (\autoref{fig:demo-ppt}D) and Affinity (\autoref{fig:demo-roundtrip}F), and synchronize in near real-time with the interface patch's slider presentation. However, it requires more discovery efforts when the states and controls are not immediately legible. For example, in the Side Views (\autoref{fig:demo-more}B), the system needs to first click on the extreme ends of the slider to discover the numerical range of the slider, and in the \`{a} la carte interface patch (\autoref{fig:demo-roundtrip}G), the system needs to discover and click on options in a dropdown, introducing delays between user's action and effect. Sometimes the pixels are inconsistent, making recovery difficult. For example, the ToolTrack interface patch (\autoref{fig:demo-more}D) counts the user's clicks on ``create sketch'' in the ribbon and in the drop-down menu separately, despite being the same feature, because the two controls have different icons, a full icon in the ribbon and a small icon in the drop-down.

Synchronization is hardest when the target application does not display the controls and values on the UI. For example, the interface patch cannot recompute the precise parameters of an object on the canvas without explicit display of such values. If the values are hidden behind layers of interface, \sys{} could periodically click and poll the values, but might introduce interruptions to the user's controls. It may be possible for \sys{} to use the pixel space to recover the object's parameters, but this would introduce measuring errors.  In these cases, APIs provide more precise and deterministic access for better modifications.

An unexpected limitation of using pixels for controls is the ``single-threaded'' nature of the OS and applications when handling user input that they expect to come from a human user. For example, macOS only provides one cursor and takes input events sequentially. As a result, the Toolglass interface patch (\autoref{fig:demo-ppt}E) cannot simultaneously accept the trackpad and the cursor input. Users cannot move the toolglass while selecting options. Additionally, it is difficult to provide compound inputs when the application does not surface multiple controls on the UI. For example, \sys{} cannot create a saliency slider that simultaneously adjusts the text colour, text size, and text weight together in PowerPoint because, without scripting access, \sys{} has to click through multiple hidden panels to modify the parameters, breaking the synchronized experience.

\subsection{Architecture and Design Rationale}

\sys{} builds upon prior systems that modify applications through creating an interface layer over the original application, including Prefab~\cite{dixonPrefabImplementingAdvanced2010}, Patina~\cite{matejkaPatinaDynamicHeatmaps2013}, and Interaction Proxies~\cite{zhangInteractionProxiesRuntime2017}. \sys{} extends this approach by combining computer-use agents with generative UI to make the modification layer more extensible. Rather than targeting a particular type of interface modifications, \sys{} supports interface patches that modify the presentation, interaction, and capabilities individually or together.

The \sys{}'s architecture design provides three benefits. First, the separation between runtime and build time retains the generalizability of build time while providing the user with a verified and deterministic modified application experience. Second, the decoupled daemon and control components in \sys{} enable retargeting a built interface patch to different OS by changing the daemon to wrap the OS APIs as \sys{}'s APIs.  Third, each interface patch is separated into the logic and specifications. This enables one interface patch implemented for one application to be easily ported to target another application without changing the interface patch's core logic.

\subsection{Limitations of \sys{} and Interface Patches}

\sys{} has several limitations in both the construction and runtime of interface patches.

At build time, creating an interface patch remains a multi-round and iterative process. A primary limitation is the agent's limited ability to verify generated interactions beyond sending basic clicks and keyboard inputs and inspecting screenshots and traces for state changes. This makes automatically testing the generated interface patches with new interaction modalities, such as gestures, pen input, or multi-device interactions, difficult. Future work could incorporate an external verification process capable of synthesizing human interface device signals to test these interactions more systematically~\cite{bighamHIDAgentToolkitEnabling2026}. Interface patch generation also takes longer than UI generation. Coding agents can generate interface patches significantly faster than a human programmer, but the individual build-test iteration can still take tens of minutes. Reusing prebuilt components in the architecture, similar interface patches implementations, and cached information about a target application reduces the construction time. However, the generation is still lacking for on-demand rapid customization due to exploration overheads when integrating the generated UI to the existing application. Reducing this latency will require faster models, as well as providing more heuristics to speed up computer-use agent explorations. 

Generating an interface patch also requires \sys{} to temporarily take control of the user's computer while it inspects and interacts with the target application, interrupting the user's normal computing activities. One possible mitigation is to recreate the user's OS and application environment in a virtual machine on the device or in the cloud, where \sys{} can generate and test the interface patch, and then distribute the resulting interface patch back on the user's device.

\sys{} introduces privacy, security, and performance concerns. First, at build time, the system has full access to operate the user's computer at build time. One potential mitigation is to sandbox \sys{} and have a secondary system firewall \sys{}'s operations and data flow. At runtime, some interface patches require \sys{} to continuously observe the application interface and aspects of the user's interaction with it. Such monitoring may expose sensitive information displayed on screen or inferred from the user's interaction traces. Running perception models locally and making most operations deterministic, like in \sys{}, reduces the amount of user data sent to external services.

Second, interface patches contain agent-generated code and may invoke external APIs and communicate over the network. As a result, generated patches may introduce vulnerabilities or inadvertently disclose user data. A practical deployment of \sys{} would therefore require stronger safeguards around generated code, including post-build inspection, sandboxing, explicit permission boundaries, and fire-walling of network access and other sensitive capabilities.

Third, continuously running interface patches, visual perception, application-state tracking, and supporting services introduce non-negligible computational overhead. \sys{} partially mitigates this cost through a shared architecture where multiple lightweight interface patch runtimes reuse common system services provided in daemon, local models, and surface rather than independently running multiple instances of them. Nevertheless, reducing runtime resource consumption remains important for deploying interface patches on more resource-constrained devices.

Finally, our current implementation requires the target application to run at the same window size as during the build time. Changes in window size, display scaling, or interface layout can therefore cause misalignment for a spatially positioned interface patch. Future versions could use the daemon's window-management and local models services to dynamically update patch geometry as the target application's window and layout change.

\subsection{Implications of Modifying Software with \sys{} and Interface Patches}

We view \sys{} as occupying an intermediate point in the lifecycle of software modification. Generative and malleable interfaces can produce interfaces in response to an immediate task or intent~\cite{caoGenerativeMalleableUser2025, vaithilingamDynaVisDynamicallySynthesized2024, xuDuetUIBidirectionalContext2026}, while conventional plugins and features delivered through software updates are engineered as stable components of an application. An interface patch sits between these two ends of the spectrum. It can originate from a local demand and be generated relatively quickly. Once built, it persists and can be repeatedly used with the target application. \sys{} does not yet make this process suitable for end-user customization, since specifying and building an interface patch still requires iteration with the coding agents. However, we demonstrate how computer use agents and generative UI can be combined to create persistent modifications to the presentation, interaction and capabilities of applications without programmatically accessing or modifying their source code. In this sense, interface patches provide a constrained form of software modification: more persistent than an ephemeral interface while more agile than conventional plugins or shipped features. In the future, \sys{} could also generate interface patches instantiated or triggered on demand, such as by an adaptive system that observes the user's interaction traces to propose modifications.

\sys{} could also serve as a design tool for higher-fidelity interaction prototyping. Rather than recreating an application as a mockup, designers could construct interface patches that operate directly with a real application with real state behaviours and data. By extending the sketch interface patch (\autoref{fig:demo-ppt}B, C), \sys{} could enable designers to combine sketch and voice to author modifications on an existing interface. This has the potential to reimagine rapid prototyping~\cite{landaySILKSketchingInterfaces1996}.

More broadly, interface patches could provide a deployment path for interaction techniques that would otherwise remain prototypes or research artifacts. \sys{} may enable revibed HCI artifacts~\cite{adarRevibingCodePapers2026} to operate directly on existing applications, allowing interaction techniques from HCI research to be deployed on any software without requiring programmatic access provided by the developer.

Beyond prototyping and research, \sys{} could become an expert customization tool. Technically proficient users could create persistent, task-specific modifications for software. This process supports rapid transitions from a feature request to tangible interface patches rather than waiting for the developer to implement the feature. Finally, the modular separation between interface patches and the OS-level daemon suggests an ecosystem of interface patches created from individual users, then shared across users, adapted to various applications and platforms, and eventually incorporated into the applications. From Interface Quilt to interface patches, the ecosystem extends the longstanding vision of customizable and community-built software~\cite{mackayPatternsSharingCustomizable1990, alvesCitizenLedPersonalizationUser2024}.

\section{Conclusion}
\label{sec:conclusion}

We presented \sys{}, a system for creating and running interface patches, persistent modifications layered over existing applications. Our demonstrations showed that interface patches can modify presentation, interaction, and capabilities across applications, through rendered-interface access alone. These results suggest that the combination of computer-use agents and generative UI can support a broader role than task automation or creating ephemeral interfaces. Together, they can reinvigorate existing software with modifications. We see interface patches as an intermediate form of malleable software, situated between ephemeral generated interfaces and fully engineered application features.

\section*{Use of Generative AI}
In accordance with the ACM Policy on Authorship, the authors disclose the use of generative AI throughout this work. 
Anthropic models (including Opus 5, Fable 5, 5.1) and the OpenAI GPT-5 family models were used across brainstorming, literature search, and prototype implementation. OpenAI's image model is used in teaser figure generation. Other local Computer Use models, UI-TARS and holo are used in system implementation. The authors reviewed and verified all generated output and take full responsibility for the Work.

\bibliographystyle{ACM-Reference-Format}
\bibliography{ref}

\end{document}